# Space Plasma Physics Science Opportunities for the Lunar Orbital Platform - Gateway

**Iannis Dandouras[1]\*, Matt G. G. T. Taylor[2], Johan De Keyser[3], Yoshifumi Futaana[4], Ruth A. Bamford[5], Graziella Branduardi-Raymont[6], Jean-Yves Chaufray[7], Dragos Constantinescu[8], Elisabetta De Angelis[9], Pierre Devoto[1], Jonathan Eastwood[10], Marius Echim[3], Philippe Garnier[1], Benjamin Grison[11], David Hercik[12], Helmut Lammer[13], André Laurens[14], François Leblanc[7], Anna Milillo[9], Rumi Nakamura[13], Lubomír Přech[15], Elias Roussos[16], Štěpán Štverák[17, 11], Julien Forest[18], Arnaud Trouche[18], Sébastien L. G. Hess[19], Jean-Charles Mateo-Vélez[19], James Carpenter[2], Josef Winter[2]**

[1]Institut de Recherche en Astrophysique et Planétologie, Université de Toulouse / CNRS / UPS / CNES, Toulouse, France

[2]ESTEC / ESA, Noordwijk, The Netherlands

[3]Royal Belgian Institute for Space Aeronomy, Brussels, Belgium

[4]Swedish Institute of Space Physics, Kiruna, Sweden

[5]RAL Space, STFC, Rutherford Appleton Laboratory, Chilton, Didcot, Oxfordshire, UK

[6]Mullard Space Science Laboratory / UCL, Holmbury St Mary, UK

[7]LATMOS (Laboratoire Atmosphères, Milieux, Observations Spatiales) / IPSL, Paris, France

[8]TU-Braunschweig, Braunschweig, Germany

[9]Institute for Space Astrophysics and Planetology / INAF, Rome, Italy

[10]Imperial College, London, UK

[11]Institute of Atmospheric Physics / CAS, Prague, Czechia

[12]scibit s.r.o., Liberec, Czechia

[13]Space Research Institute / Austrian Academy of Sciences, Graz, Austria

[14]CNES, Toulouse, France

[15]Charles University, Prague, Czechia

[16]Max Planck Institute for Solar System Research, Göttingen, Germany

[17]Astronomical Institute / CAS, Prague, Czechia

[18]Artenum, Ramonville Saint-Agne, France

[19]ONERA, Toulouse, France

**\* Correspondence:**
Corresponding Author
Iannis.Dandouras@irap.omp.eu

**Abstract**

The Lunar Orbital Platform - Gateway (LOP - Gateway, or simply Gateway) is a crewed platform that will be assembled and operated in the vicinity of the Moon by NASA and international partner organizations, including ESA, starting from the mid-2020s. It will offer new opportunities for fundamental and applied scientific research. The Moon is a unique location to study the deep space plasma environment. Moreover, the lunar surface and the surface-bounded exosphere are interacting with this environment, constituting a complex multi-scale interacting system. This paper examines the opportunities provided by externally mounted payloads on the Gateway in the field of space plasma physics, heliophysics and space weather, but also examines the impact of the space environment on an inhabited platform in the vicinity of the Moon. It then presents the conceptual design of a model payload, required to perform these space plasma measurements and observations. It results that the Gateway is very well-suited for space plasma physics research. It allows a series of scientific objectives with a multi-disciplinary dimension to be addressed.

## 1    Introduction

The Moon is a unique location to study the deep space plasma environment. During most part of its orbit around the Earth the Moon is directly exposed to the solar wind. Due to the absence of a substantial intrinsic magnetic field and of a collisional atmosphere, solar wind and solar energetic particles (SEPs) arrive almost without any deviation or absorption and impact directly on its surface, interacting with the lunar regolith and the tenuous lunar exosphere (e.g. Geiss et al., 2004; Futaana et al., 2018). The same phenomenon occurs also with the galactic cosmic rays (GCRs), which present fluxes and energy spectra typical of interplanetary space (e.g. Sohn et al., 2014). Downstream from the Moon, a structured plasma umbra and penumbra region is formed, characterized by the gradual decrease of the ion and electron densities (Bosqued et al., 1996; Nishino et al., 2010). The Moon's vicinity is an ideal environment to study galactic cosmic rays, solar wind and solar energetic particles. This environment is typical of deep space (Plainaki et al., 2016), apart from the fact that the Moon itself forms an obstruction to the GCRs.

During 5 – 6 days every orbit, however, the Moon crosses the tail of the terrestrial magnetosphere (**Figure 1**). It is then exposed not to the solar wind but to the terrestrial magnetotail plasma environment, offering the possibility to study in-situ magnetotail dynamics and its dependence on solar and geomagnetic activity (e.g. Kallio and Facskó, 2015; Kallio et al., 2019). Phenomena such as plasmoids released from the near-Earth magnetotail and propagating anti-Sunward, bursty bulk flows (BBFs), energetic particle bursts, plasma waves, magnetic reconnection and plasma sheet dynamics can thus be studied in-situ (e.g. Parks et al., 2001; Nakamura, 2006; Taylor et al., 2006; Nagai et al., 2009; Du et al., 2011; Cao et al., 2013; Grigorenko et al., 2019; Sitnov et al., 2019; Kronberg et al., 2021).

The Moon is then also very well situated to study atmospheric escape from the Earth into space (Lammer et al., 2008; Harnett et al., 2013; Wei et al., 2020; Dandouras, 2021; André et al., 2021; Wang et al., 2021), in the form of heavy ions upwelling from the terrestrial ionosphere and transported and lost into the deep magnetotail. The wealth of data supplied from the THEMIS-ARTEMIS and from the Kaguya (SELENE) spacecraft confirmed the observation of such ions, of terrestrial origin, in the lunar environment (Poppe et al., 2016; Terada et al., 2017). The THEMIS-ARTEMIS data, however, did not include the crucial information on the plasma composition (Angelopoulos, 2011), and the Kaguya plasma measurements where limited to a less than two-year mission and to low-energy plasma (Saito et al., 2010). Far magnetotail studies performed by the Geotail spacecraft supplied key information on the dynamics of ion beams streaming downtail (Christon et al., 1994, 2020; Seki et al., 1998), but lacked the ion composition measurements at low energies (below ~10 keV).

When the Moon gets again outside of the magnetotail, terrestrial magnetosphere dynamics can be monitored through remote sensing, using a variety of magnetospheric imaging techniques. These include Energetic Neutral Atom (ENA) imaging, which conveys information on the interaction between energetic ions and the terrestrial exosphere (e.g. C:son Brandt et al., 2002; Vallat et al., 2004), solar wind charge exchange X-rays imaging of the interaction between the solar wind / magnetosheath plasma and the terrestrial exosphere (Branduardi-Raymont et al., 2012; Sibeck et al., 2018), plasmasphere EUV imaging (Sandel et al., 2003), or exosphere Lyman-α imaging (e.g. Zoennchen et al., 2017).

But, most important, the lunar environment offers a unique opportunity to study the Moon surface-bounded exosphere (**Figure 2**), its production mechanisms, its dynamics, its interaction with the solar wind and with the terrestrial magnetotail plasma, and its escape into space (Potter et al., 2000; Wurz et al., 2007, 2022; Futaana et al., 2008; Leblanc and Chaufray, 2011, Lammer et al., 2022). The LADEE (Lunar Atmosphere and Dust Environment Explorer) and LRO (Lunar Reconnaissance Orbiter) observations have provided a glimpse of the complexity of the lunar exosphere and of the associated physical mechanisms (Stern et al., 2013; Elphic et al., 2014; Benna et al., 2015; Hodges, 2016; Hurley et al., 2016).

The lunar surface offers also exciting possibilities for studying energetic ion implantation in the lunar regolith (Ozima et al., 2005; Ireland et al., 2006), albedo energetic particles produced through the interaction of SEPs and GCRs with the regolith (Schwadron et al., 2016, 2018; Zhang et al., 2020), solar wind ion implantation or neutralization and reflection from the lunar regolith (Futaana et al., 2006, 2012; Vorburger et al., 2015; Tucker et al., 2019), formation of hydrogen bearing molecules (McCord et al., 2011; Stern et al., 2013; McLain et al. 2021) possibly including water (Schörghofer et al., 2021), solar wind interaction with crustal magnetic anomalies (Poppe et al., 2015; Bamford et al., 2016), lunar pickup ion generation (Poppe et al., 2012a; Wang et al., 2011), or lunar surface electrostatic charging and dust levitation (Stubbs et al., 2007; Hess et al., 2015; Popel et al., 2018), just to mention few examples.

The analysis of implanted particles on the lunar surface, that originated from the Earth's atmosphere, will also reveal some knowledge of Earth's early atmosphere (Marty et al., 2003; Ozima et al., 2005; Lammer et al., 2018, 2022). It is expected that early Earth's atmosphere experienced strong escape of hydrogen, oxygen and carbon, that originated from the dissociation of water and methane molecules, and of nitrogen due to the increased EUV flux from the young Sun (Lammer et al., 2018; Zahnle et al., 2019; Gebauer et al., 2020; Kislyakova et al., 2020; Johnstone et al., 2021). As suggested by Marty et al. (2003), nitrogen originating from the early Earth was implanted on the lunar surface. This is based on the strong variations of N, He, Ne and Ar noble gas isotope implantations into the regolith, of up to 30 % (Ozima et al., 2005). According to Marty et al. (2003) and Ozima et al. (2005) these enhancements cannot be explained as due to solar wind implantation alone.

The Moon is also an ideal test case for studying planetary surface weathering resulting from the exposure to energetic particles, i.e. surface - energetic particle interactions (e.g. Hapke, 2001; Pieters and Noble, 2016; Nénon and Poppe, 2020). Given that the Moon is irradiated by GCRs quasi-uniformly, any differences in the resulting interaction, including the emitted albedo particles, point to the variable properties, physical or chemical, of the surface (Schwadron et al., 2016, 2018). From the perspective of the Gateway, surface - GCR interactions can be mainly probed through the albedo particles.

Following the legacy of the Apollo missions and of the more recent missions to the Moon (THEMIS-ARTEMIS, Kaguya, LADEE, LRO, Chandrayaan, Chang'e, etc.), a series of lunar missions is in preparation, or already operating, building on their outstanding heritage (Dandouras et al., 2020a).

The Lunar Orbital Platform - Gateway (LOP - Gateway, or simply Gateway) is a crewed platform that will be assembled and operated in the vicinity of the Moon by NASA and international partner organizations, including ESA. Launch of the first modules will start in the mid-2020s (Phase 1), and it will continue with the launch and assembly of additional modules during the late 2020s (Phase 2). The Gateway will provide support for all lunar activities, including the Artemis program to return humans to the Moon (Artemis III Science Definition Team Report, 2020). It will also offer new opportunities for fundamental and applied scientific research (Carpenter et al., 2018; Dandouras et al., 2020a).

In preparation of its scientific payload, ESA set up international science teams to prepare and to support the definition of payload studies, including a topical team in the field of space plasma physics. In the first part of this article (sections 2 and 3) we report on the outcome of this topical team, which was entitled "Space Plasma Physics Science Opportunities for the Lunar Orbital Platform - Gateway". This part focuses on the science objectives identified by the topical team (section 2), and on the corresponding instrumentation required to address them (section 3). In the second part (section 4) we present a conceptual design study for a "Space Plasma Physics Payload Package onboard the Gateway" (SP4GATEWAY) we undertook for ESA, addressing these objectives and compatible with the technical requirements.

## 2    Specific Objectives and Goals

The "Space Plasma Physics Science Opportunities for the Lunar Orbital Platform-Gateway" topical team, set up by ESA in 2019, brought together the key expertise required for defining the space plasma parameters to measure from lunar orbit, and the appropriate instrumentation required to perform these observations. The science objectives that were identified include:

### 2.1    Monitoring the solar wind and the lunar energetic particle environment

Due to the absence of a substantial intrinsic magnetic field and of a collisional atmosphere, the Moon is directly exposed to:

– Solar Wind: ~keV particles
– Solar Energetic Particles (SEPs): ~MeV particles
– Galactic Cosmic Rays (GCRs): ~GeV particles

The monitoring of the solar wind (e.g. von Steiger, 2008), at the lunar environment, aims to evaluate its role as a driver for the dynamics of the terrestrial and the lunar exospheres, of the dynamics of the terrestrial magnetosphere, and of the lunar surface sputtering and charging.

The monitoring and characterization of the SEPs and GCRs, at lunar orbit, aims to evaluate the radiation environment of the Moon and also the role of SEPs and GCRs as lunar surface sputtering sources. Since the Moon does not have a substantial magnetic field it is possible, with an appropriate particle detector, to measure the low energy part of the GCR spectrum (< 1 GeV) with high precision.

This offers an advantage with respect to low-Earth orbits, where most of the advanced GCR observatories like PAMELA and AMS-02 are located, where this low energy part is filtered out by the Earth's magnetosphere.

Typical SEP proton intensities, measured during a solar event, are shown in **Figure 3** (adapted from Quinn et al., 2017). Some of the SEP protons (~MeV energy range) can interact, in the high solar corona, with partially stripped coronal ions, charge exchange with them and produce ~MeV ENAs (Energetic Neutral Atoms) (Mewaldt et al., 2009).

GCR Hydrogen and Oxygen nuclei fluxes are shown in **Figure 4**, presenting a clear solar cycle modulation (adapted from Mrigakshi et al., 2012). The interaction of these SEPs and GCRs with the lunar regolith produces albedo energetic particles, resolvable with current instruments up to a few ~100 MeV, and with fluxes that are sensitive to the regolith hydration (Looper et al., 2013; Schwadron et al., 2016; Zaman et al., 2022), cf. **Figure 5**. The separation of the pristine energetic particle fluxes from the albedo energetic particles (e.g. by zenith centered / nadir centered looking directions respectively) appears thus as a requirement, in order to provide information on the deep space SEP and GCR environment and on the interaction of the lunar regolith with this environment.

## 2.2    Monitoring the terrestrial magnetosphere and exosphere

When the Moon is within the terrestrial magnetotail, in-situ measurements of the plasma sheet and plasma sheet boundary layer dynamics are enabled. These consist of magnetic field and energetic ion and electron monitoring, including the measurement of energetic ions of terrestrial origin streaming downtail.

The evolution of the flux of $O^+$ downtail streaming beams, as a function of the tailward distance from the Earth, is shown in **Figure 6** (from Seki et al., 1998). During high geomagnetic activity conditions these beams include heavy atomic and molecular ions (Christon et al., 1994, 2020). Closer to the Moon, $O^+$ downtail streaming beams have been observed by the Kaguya Lunar Orbiter (Terada et al., 2017). The spectral characteristics of these streaming $O^+$ ions show a clear distinction between the $O^+$ ions of lunar origin (few 10 eV to ~100 eV) and the terrestrial magnetospheric $O^+$ ions (few keV), cf. **Figure 7** (adapted from Terada et al., 2017). Particle tracing simulations performed by Harnett et al. (2013) using a 3D multi-fluid model, and by Poppe et al. (2016) using the MHD Open Global Geospace Circulation Model, show how heavy ions, originating from the Earth's inner magnetosphere, can be ejected downtail during high geomagnetic activity events, reaching energies of several keV to several 10 keV at lunar distances, cf. **Figure 8**.

When the Moon is outside of the magnetotail, terrestrial magnetosphere dynamics and response to solar wind conditions can be monitored through remote sensing. This includes:

–   Ring current and near-Earth plasma sheet monitoring, by imaging of the ENAs produced by charge exchange between the plasma sheet or ring current energetic ions (few ~keV to few ~10 keV) with the geocorona neutral hydrogen atoms, e.g. Brandt et al. (2004), Vallat et al. (2004), Goldstein et al. (2022).
–   Magnetopause and cusps monitoring by detecting and imaging the SWCX (solar wind charge exchange) soft X-rays produced by charge exchange between highly-charged heavy ions, originating from the solar wind, and the exospheric neutral atoms, e.g. Branduardi-Raymont et al. (2012, 2021), Sibeck et al. (2018).
–   Plasmasphere imaging, by resonant scattering of the solar EUV (30.4 nm) by the plasmaspheric $He^+$ ions, e.g. Sandel et al. (2003), Darrouzet et al. (2008), He et al. (2016).

- Geocorona imaging at Lyman-α (121.6 nm), e.g. Rairden et al. (1986), Zoennchen et al. (2017, 2022).

## 2.3 Monitoring the Moon's surface-bounded exosphere

The Moon surface-bounded exosphere constitutes a complex multi-scale system (**Figure 2**), characterised by its interactions with the solar radiation, the solar wind and terrestrial magnetotail plasma, the meteoritic flux, dust, and the regolith (Futaana et al., 2018). The low number densities of this very tenuous atmosphere, particularly of the minority species, and the complexity and multiplicity of the source and loss mechanisms have resulted in a poor understanding of it (Wurz et al., 2007, 2022; Poppe et al., 2022). **Figure 9** provides the altitude density profiles of the major species, separately for the atoms and molecules released thermally from the regolith and for the atoms released through sputtering. In addition to atomic and molecular hydrogen, O, OH, $CH_4$, noble gases (He, Ne, Ar, Xe), metallic atoms (Na, K, Mg, Al) and other elements populate the lunar exosphere (Leblanc and Chaufray, 2011; Benna et al., 2015; Grava et al., 2015, 2016, 2021; Halekas et al., 2015; Hodges, 2016; Hurley and Benna, 2018; Leblanc et al., 2022; Wurz et al., 2022).

These neutral exospheric atoms and molecules can be subsequently ionized by the solar UV radiation and generate pickup ions. These ions are promptly accelerated from their birthplace by the ambient electric field **E** and drift across the magnetic field **B**. The unique orbital characteristics of the pickup ions (cycloidal motion consisting of a combination of **E** × **B** drift and a gyration around **B**) make it possible to infer important details about their sources (Hartle and Killen, 2006). Such lunar pickup ions have been detected in the terrestrial magnetotail lobes (Poppe et al., 2012a) and in the solar wind (Wang et al., 2011).

As the measurements performed onboard the LADEE and LRO spacecraft have shown, the lunar exosphere can be monitored, either by in-situ measurements, using a neutral mass spectrometer, or by remote sensing using a UV spectrometer (Chin et al., 2007; Elphic et al., 2014).

Other techniques for studying the lunar exosphere are:

- Remote sensing of the lunar exosphere by detecting and imaging the ENAs produced by charge exchange interactions between the solar wind protons and the exospheric neutral atoms (Futaana et al., 2008). The energies of these ENAs are comparable to the energies of the parent solar wind protons, i.e. of the order of ~keV.
- Remote sensing of the lunar exosphere by detecting and imaging the SWCX soft X-rays produced by charge exchange between highly-charged heavy solar wind ions and the exospheric neutral atoms (Robertson et al., 2009).
- In-situ measurement of freshly ionized pickup ions, originating from the lunar exosphere neutral species. (Hartle and Killen, 2006; Yokota et al., 2009; Wang et al., 2011; Poppe et al., 2022). At high altitudes above the lunar surface, as those of the Gateway orbit (cf. section 4.1), this method can provide higher sensitivity in the detection of low number density species than the direct sampling of the parent neutrals (Halekas et al., 2015; Poppe et al., 2022).

## 2.4 Monitoring the interaction of the solar wind with the Moon's surface

Solar wind protons, arriving at the Moon's surface, can be absorbed, or scattered, or can remove another atom from the lunar regolith by sputtering or desorption (Wieser et al., 2009; McComas et al., 2009a; Futaana et al., 2012). It results that a large fraction of the solar wind protons, up to 20%, is reflected back to space as neutral hydrogen atoms (ENAs). It is noteworthy that backscattering of neutralized solar wind protons occurs not only when the Moon is in the pristine solar wind, but also when the Moon enters into the terrestrial magnetosheath and is then exposed to the shocked and thermalized solar wind (Allegrini et al., 2013).

**Figure 10** shows typical energy spectra of the reflected hydrogen ENAs, compared to the parent solar wind protons energy spectra. As shown, the flux of the reflected ENAs closely follows the variations of the flux of the parent proton population. The energies of these ENAs are however a fraction of the parent solar wind protons.

Since the solar wind proton trajectories are modulated by the surface electrostatic potential and by the eventual local magnetic field anomalies (cf. **Figure 11**), the detection and imaging of these reflected ENAs provides a tool to investigate the lunar surface electric and magnetic fields (Futaana et al., 2013; Vorburger et al., 2013, 2015, 2016; Bamford et al., 2016). Local crustal magnetic anomalies (or "swirls") constitute "mini-magnetospheres", shielding locally the lunar regolith from the solar wind protons and from the resulting space weathering (Wieser et al., 2010; Wang et al., 2012; Deca et al., 2015; Glotch et al., 2015; Hemingway et al., 2015; Poppe et al., 2015; Pieters and Noble, 2016; Hemingway and Tikoo, 2018).

The solar wind protons that do not scatter back, but are absorbed in the lunar regolith (top 20 – 30 nm of the lunar grains), diffuse within the regolith. They can then interact with the oxygen atoms in the regolith oxides and form OH (McCord et al., 2011; Farrell et al., 2017; Tucker et al., 2019; McLain et al., 2021). These solar wind-produced hydroxyl radicals contribute to the formation and release of molecular water, and thus to a solar wind-induced water cycle on the Moon (Crider and Vondrak, 2003; Liu et al., 2012; Futaana et al., 2018; Jones et al., 2018; Honniball et al., 2021).

The exposure of the lunar surface to the solar radiation and to the flux of charged particles results also in an electrostatic surface charging. An electric potential thus develops between the lunar surface and the ambient plasma, which manifests itself in a near-surface plasma sheath with a scale height of the order of the Debye length (Halekas et al., 2011; Stubbs et al., 2013; Burinskaya, 2015; Harada et al., 2017). This near-surface electric field becomes very complex and highly variable in the vicinity of the terminator, with the surface polarity changing from mostly positive (few 10 V) on the dayside, due to photoelectron emission, to highly negative (of the order of the ambient electron temperature, i.e. up to several -100 V) on the nightside, and in the trailing lunar wake region (Farrell et al., 2007). Local surface topography is also a factor contributing to a complex near-surface electrostatic and plasma environment, particularly in the vicinity of permanently-shadowed craters (Poppe et al., 2012b; Nénon and Poppe, 2021). As the THEMIS-ARTEMIS observations have shown, the lunar surface charging can be remotely sensed from a Moon orbiting spacecraft, even several 1000 km away from the lunar surface, through the shifted energy spectra of the detected plasma particles when the spacecraft crosses magnetic field lines connected to the lunar surface (Halekas et al., 2011).

Dust is another component of the lunar plasma environment (Stubbs et al., 2007; Grün et al., 2011; Horányi et al., 2015; Popel et al., 2018, 2022). Dust grains on (or near) the lunar surface can either be ejected from the regolith, due to the impact of interplanetary micrometeoroids, or be electrostatically levitated due to grain charging, as discussed in the previous paragraph. This creates a dusty plasma system consisting of neutrals of the lunar exosphere, solar-wind ions and electrons, ions and electrons of the Earth's magnetotail (when the Moon gets inside the terrestrial magnetotail), photoelectrons formed due to the interaction of the solar radiation with the lunar surface, and charged dust grains flying over the lunar surface.

## 3 Measurement Requirements

Following the identification of the scientific objectives in the field of space plasma physics, that can be addressed using instrumentation onboard the Lunar Orbital Platform - Gateway (cf. section 2), the ESA topical team identified the physical parameters needed to be measured in order to address these objectives, and the corresponding instrumentation required to perform these observations. The topical team addressed thus the following two questions (Dandouras et al., 2020b):

–   What plasma physics science questions can be addressed in the vicinity of the Lunar Orbital Platform - Gateway?
–   What are the instrument / payload requirements to achieve such science?

It identified measurements that can be performed either directly from the Gateway platform (3 200 × 70 000 km altitude lunar orbit), or from instrumented cubesats that could be released from the platform and placed into lower lunar orbits, or directly from the Moon surface. Here we will focus on the measurements that can be performed by instrumentation mounted onboard the Gateway, and which can be either in-situ measurements or remote sensing observations, and then we briefly mention the other two possibilities. Space plasma physics measurements that could be performed directly from the Moon surface will be the object of a dedicated forthcoming paper.

**Table 1** provides an overview of the physical parameters / observables identified, in the field of space plasma physics, that can that be monitored by instrumentation onboard the Gateway.

**Tables 2 and 3** are for the observations that could be performed, on a longer term, from lower lunar orbits and from the Moon's surface, respectively.

**Table 4** focuses then on the science questions that can be addressed from instrumentation onboard the Gateway, and shows how each science objective, identified by the topical team, translates into a measurement requirement, and then to the corresponding instrument / payload requirement.

Additional objectives that could be eventually addressed by remote sensing instrumentation onboard the Gateway, and could point to targets of opportunity, include: aurora imaging, heliosphere imaging (through the ENA imager), and lunar surface imaging (e.g. meteor impact flashes).

## 4 Conceptual design for a Space Plasma Physics Payload Package onboard the Gateway

Following the work of the topical team, and the identification of the measurement requirements, ESA issued an Invitation to Tender for a "Deep Space Gateway Plasma Physics Payload Conceptual Design" (ESA AO/1-9789/19/NL/FC). In response to it we proposed to ESA, were selected and then undertook a conceptual design study for a "Space Plasma Physics Payload Package onboard the Gateway" (SP4GATEWAY), addressing these objectives while being compatible with the technical requirements.

The Gateway modules that are best-suited for hosting the in-situ measurement plasma instruments were first identified, following a simulation we performed of the interaction between the Gateway and its plasma environment (section 4.2). The proposed model payload, and its accommodation on the Gateway modules, are presented in sections 4.3 and 4.4 respectively. The fields-of-view (FOVs) of the remote sensing instruments, as projected on the sky and on the celestial objects, were then analyzed by simulating their evolution along the Gateway orbit (section 4.5).

## 4.1 Gateway configuration, orbit and attitude

The Gateway will evolve during its lifetime, different modules being added during the successive phases of the project. For the purpose of this study we considered a typical "Gateway Phase 2" configuration, with the Orion spacecraft attached, shown in **Figure 12**.

The Gateway orbit will be a Near Rectilinear Halo Orbit (NRHO) around the Moon, with periapsis × apoapsis altitudes 3 200 × 70 000 km (Whitley and Martinez, 2016). The orbital period is ~6.5 (Earth) days, and the orbital inclination ~90°. The periapsis will be above the north pole of the Moon. This orbit provides constant Earth visibility (9:2 resonance with the lunar synodic period).

The Gateway attitude will be with the +X axis (longitudinal axis, cf. **Figure 12**) pointed towards the Sun. The +Z axis will be normal to the Moon orbit plane, pointing southwards. The pointing accuracy requirement is that Orion remains in a tail-to-Sun attitude ±20°, i.e. the +X axis has a ±20° pointing accuracy.

## 4.2 Simulation of the Gateway plasma environment

The simulation of the interaction between the Gateway and the plasma environment was performed by ONERA and the Artenum company (Hess et al., 2020). A 3D mesh model with approximately 64 000 elements was developed to represent the Gateway, and the properties of the surface materials of the different Gateway modules were taken into account. The SPIS (Spacecraft Plasma Interaction System) software tool was then used to simulate the Gateway interaction with its ambient plasma environment. This open-source software, available at https://www.spis.org, computes the potential at the surface of a spacecraft according to its exchange of charges with the space plasma, i.e. the collection of charge from the plasma and the re-emission of photoelectrons and of secondary electrons due to impacting energetic particles. It also simulates the perturbation induced by this electrostatic charging on the natural plasma. This software was further developed to simulate the charging of the regolith and the motion of lunar dust particles (Hess et al., 2015) and to simulate the perturbation of the measurements by plasma instruments due to the charging (Sarrailh et al., 2015).

Two cases were simulated, that correspond to the two situations that will be typically encountered:

1. Gateway in the solar wind (most frequent case, cf. section 1).
   Typical solar wind conditions considered were:
   solar wind density: 7 cm$^{-3}$
   solar wind velocity: 450 km/s
   ion and electron temperatures: 10 eV

2. Gateway in the terrestrial magnetotail (5 – 6 days per lunar orbit).
   The plasma environment considered, corresponding to active geomagnetic activity conditions (conditions producing downtail plasma streaming, cf. section 2.2), was:
   plasma density: 2.01 cm$^{-3}$
   H$^+$ density: 2 cm$^{-3}$
   O$^+$ density: 0.01 cm$^{-3}$
   plasma streaming velocity: 250 km/s (away from the Earth)
   ion temperature: 200 eV
   electron temperature: 15 eV

In each case both a nominal Gateway attitude (Gateway major axis aligned to the solar direction, cf. section 4.1) and an extreme attitude excursion, with the Gateway major axis tilted by 20° with respect to the solar direction, were considered. The simulation runs generated, for each case, maps of the electrostatic potential (volume values in the Gateway environment and surface values on the Gateway modules), and maps of the density values of $H^+$, $O^+$, photoelectrons and secondary electrons. The details are given in the report by Hess et al. (2020).

The main results of this study are:

### 4.2.1 Gateway in the solar wind

The volume electrostatic potential distribution, when the Gateway is in the solar wind and the major axis of the station is aligned to the solar direction (nominal attitude), following 600 s of interaction time, is shown in **Figure 13**.

The Gateway structure gets to a 3.5 V equilibrium potential, while the major part of the solar panels goes to a 10 V potential on the Sun facing side and -46 V on the rear side (**Figure 13A**). The PPE (Power and Propulsion Element), bearing the two main solar panels, is thus inappropriate for space plasmas instrumentation for low-energy plasmas. The wake effect, due to the solar wind flow, is particularly visible behind the main solar panels, whereas in the front modules of the station the potential perturbation appears to be moderate. To highlight the potential values away from the solar panels (i.e. where the plasma instruments should be mounted), the surface and volume potentials are plotted also in a scale saturated between +5 V and -3 V (**Figure 13B**). As shown there, the thickness of the sheath formed by the plasma flow around the Gateway, on the station parts exposed to the solar wind and away from the solar panels, is typically ~1.8 m and the electrostatic potential perturbation is moderate (a few volts). This implies that the effect on the ion and electron measurements will be very moderate, and only the lowest energy particles (< ~100 eV) will be affected. Solar wind ions, which have energies of typically ~1 keV, will be almost not affected. It implies also that a boom of ~2 – 3 m length is adequate for placing sensors as a magnetometer and a wave antenna outside of the sheath.

The ambient proton density around the Gateway is shown in **Figures 13C** and **13D**. The left panel (**Figure 13C**) corresponds to the nominal Gateway attitude (Gateway major axis aligned to the solar direction), whereas the right panel (**Figure 13D**) corresponds to an extreme attitude excursion of 20° with respect to the solar direction. Note, in both cases, the plasma wake downstream of the station. The tilted axis simulations show a small asymmetry between the illuminated and the shadowed sides of the Gateway and, as expected, a tilted plasma wake.

### 4.2.2 Gateway in the terrestrial magnetotail

Here the Gateway is exposed to the terrestrial plasma sheet / magnetosheath plasma streaming downtail. The volume electrostatic potential distribution, under these conditions and when the major axis of the station is aligned to the solar direction (nominal attitude), following 400 s of interaction time, is shown in **Figure 14A**. The surface equilibrium potential here is 6.5 V, while the major part of the solar panels goes to a 13 V on the Sun facing side and -31 V on the rear side. Due to the lower density of the ambient plasma, the sheath forming around the station is more extended, but the potential barrier is weaker (-1.1 V) and more isotropic, compared to the solar wind case. However, the overall results are not very different and the conclusions made in the solar wind case apply also here. **Figure 14B** shows the emitted photoelectron density.

The ambient $H^+$ and $O^+$ ion densities (terrestrial ions streaming downtail during active geomagnetic conditions, cf. section 2.2) are shown in **Figures 14C** and **14D** respectively. The $O^+$ density distribution shows here a high similarity with the proton density distribution in the solar wind, presenting a very clear wake effect due to the higher ion mass.

### 4.2.3 Gateway - plasma environment interaction: synthesis

The interaction of the Gateway with its plasma environment has been simulated for the two cases that will be encountered: Gateway in the solar wind and Gateway in the terrestrial magnetotail. In both cases the surface potential of the Gateway away from the solar panels is moderate (3.5 V in the solar wind and 6.5 V in the magnetotail). A sheath is formed by the plasma flow around the Gateway, which for the solar wind case has a thickness of ~1.8 m when the Gateway is aligned to the solar direction. However, when the Gateway major axis (X-axis) is tilted by 20° with respect to the solar direction, which corresponds to an extreme excursion from the nominal attitude, this plasma sheath becomes asymmetrical and much thicker in the "shadowed" side.

These results are very encouraging, because they allow to identify the Gateway modules on which the perturbation of the natural plasma environment by the Gateway will be minimal, and are thus well-suited for placing the plasma instruments. **Figure 15** shows the positions identified for instrument mounting in a color code, from green (most favorable) to red (least favorable positions).

The US Habitat and the International Habitat present small surface charging, are surrounded by a thin plasma sheath and do not suffer from any plasma wake effect. They are thus suitable for placing the plasma instruments sensitive to electrostatic charging, as the magnetospheric ion and electron spectrometers (green / light green markers in **Figure 15**).

However, these positions on the cylindrical surfaces of the US Habitat and of the International Habitat are tangent to the solar wind flow. When the Gateway is tilted with respect to the solar direction, the solar wind flow is detached and thus not measurable from these positions (cf. **Figure 13D**). Solar wind measurements require, not only limited (less than ~10 V) surface charging and absence of local plasma wake effects, but also a direct "face exposure" to the solar wind. The +X side of the Logistics Module (lower light green marker in **Figure 15**) is thus the most suitable position for the solar wind instruments.

Concerning the wave and field instruments, their positioning on ~2 – 3m booms, on the "green / light green markers", allows having them outside of the plasma sheath.

The remaining positions can be used for energetic particle and magnetospheric imaging instruments, which are not sensitive to plasma charging effects.

The least favorable positions for placing plasma instruments (positions to avoid) are the PPE (Power and Propulsion Element) and the close to it HALO (Habitation and Logistics Outpost), cf. red markers in **Figure 15**, due to the large solar panels and associated circuitry, their "downstream" positioning (with respect to the solar wind flow), the high surface charging, and the proximity to the ion propulsion engine.

### 4.3 Model payload

In order to address the scientific objectives identified by the topical team, we first defined a model payload, consisting of a suite of instruments corresponding to the requirements shown in **Table 4** (cf. section 3). These measurement instruments are largely based either on existing flight-proven instruments, adapted here for the lunar plasma environment, or on tested and validated laboratory prototypes (TRL (Technology Readiness Level) ≥ 5). **Table 5** lists these instruments and provides an overview of their main characteristics. The detailed description of the characteristics of the instruments is given in a series of three ESA reports, corresponding respectively to a Requirements Inventory (De Keyser et al., 2020), Conceptual Design Report (Devoto and Dandouras, 2020), and Programmatic Assessment (Futaana et al., 2020). Here we present their principal characteristics.

### 4.3.1 cMAGF: 3-axis Fluxgate Magnetometer

This instrument will provide the ambient vector magnetic field (in solar wind, terrestrial magnetotail, Moon vicinity, lunar wake, etc.). The proposed magnetometers package consists of three different types of units. The main one has three pieces of boom-mounted 3-axis fluxgate magnetometers, on two ~3 m retractable booms: one sensor at the tip of each boom, and a third sensor at the common root of the two booms. This allows having two main sensors outside the plasma sheath formed by the plasma flow around the Gateway (cf. section 4.2), while the boom-root sensor provides the possibility for removing eventual Gateway-induced perturbations by using the gradiometer technique. The presence of two boom-tip mounted sensors, on two booms, provides for further corrections for eventual perturbations. These three units will provide the main measurements while supporting the cleaning and processing of the measured data.

In order to monitor the perturbations from the station in more detail, several single magnetometer sensors will be mounted on various places directly on the station (Constantinescu et al., 2020). Additionally, we propose two current monitors, monitoring the currents flowing from the solar panels, which are expected to contribute the largest magnetic field perturbations.

The proposed accommodation on the Gateway of this magnetometer package, and the corresponding CAD figures, are shown as also for the other instruments in section 4.4.

Fluxgate magnetometers benefit from a strong heritage, as such instruments have flown on several space missions, including Cluster (Balogh et al., 2001), Cassini (Dougherty et al., 2004), THEMIS-ARTEMIS (Auster et al., 2008), BepiColombo (Glassmeier et al., 2010), etc.

### 4.3.2 cSWIS: Solar Wind Ion Spectrometer

The cSWIS instrument is a solar wind ion spectrometer that will determine the velocity distribution functions (VDFs) of the solar wind ions and will provide the solar wind density, velocity and temperature.

A top-hat electrostatic analyzer instrument is considered, covering the 0.1 – 40 keV/e energy range and having a field-of-view (FOV) of 96° × 48° aligned with the solar wind arrival direction: 96° angular range in azimuth (+24° to -72° in the ecliptic plane, as the Gateway points to the Sun but it may sometimes drift away from this direction, after which it catches up) and 48° in elevation (between -24° to +24°). The spatial resolution is 3° in both azimuth and elevation, and the energy resolution is $\Delta E/E = 8\%$. In order to achieve high VDF acquisition cadence, we propose to use solar wind beam tracking, along the lines of the Cold Solar Wind (CSW) instrument (Cara et al., 2017; De Keyser et al., 2018), that was designed for the THOR (Turbulent Heating ObserveR) mission which was proposed to ESA as a medium-class M4 mission.

### 4.3.3  cSWFC: Solar Wind Faraday Cup

The cSWFC instrument will be used to determine the solar wind density, velocity and relative alpha-particle content, based on simultaneous measurements of the collector currents provided by six identical Faraday cups.

The energy of incoming ions is determined by the high voltages applied onto the control grids. The Faraday cups are organized into three units, each of them containing two cups. One unit serves for the determination of the total ion flux vector, the second unit uses high voltages applied on the control grids and provides two points of energy distribution that are used for the determination of the proton velocity and temperature in the Maxwellian approximation. The last unit serves for the measurement of the 1D velocity distribution (integral distribution) of protons and alpha particles. Each of the six Faraday cups will have a 45° × 45° FOV which, as for the cSWIS instrument, will be aligned with the solar wind arrival direction. The energy resolution is 1% (< 50 eV).

The proposed cSWFC instrument is based on the BMSW (Bright Monitor of the Solar Wind) Faraday cup instrument, that flew onboard the Spektr-R mission (Šafránková et al., 2013).

### 4.3.4  cMISP: Magnetospheric Ion Spectrometer

cMISP is a mass-discriminating ion spectrometer, that determines the velocity distribution functions of the ambient plasma ions: terrestrial magnetosphere ions, lunar exosphere pickup ions and solar wind ions.

The proposed instrument is a time-of-flight ion mass spectrometer capable of obtaining ion distributions (about 10 eV/e to 40 keV/e) with a high-resolution mass-per-charge composition determination ($m/\Delta m > 15$). Ions are selected as a function of their $E/q$ (energy per charge) ratio, by sweeping the high voltage applied between the two hemispheres of a rotationally symmetric toroidal electrostatic analyzer (360° ×5° instantaneous FOV). Then they go through a post-acceleration of about 5 kV and they subsequently enter into the time-of-flight (TOF) section, where the velocity of the incoming ions is measured, which allows then the calculation of their $m/q$ (mass per charge) ratio. A specially designed thin microchannel plate (MCP), through which the ions pass, is used as a conversion surface for the production the "start" TOF signal secondary electrons. The "stop" TOF signal is provided by the ion detection on another MCP. The instrument provides for a $\Delta E/E$ ~7 % energy resolution and a 22.5° angular resolution.

cMISP is based on the MIMS (MCP Ion Mass Spectrometer) instrument, that was designed for the ESCAPE (European SpaceCraft for the study of Atmospheric Particle Escape) mission, which was proposed to ESA as a medium-class M5 mission (Dandouras et al., 2018). MIMS in its turn was based on a successfully tested prototype developed at IRAP (Devoto et al. 2008). MIMS is an evolution of the CIS-CODIF instrument, flying onboard Cluster (Rème et al., 2001), but with higher mass resolution.

Since MIMS was designed for a spinning spacecraft, where it would take advantage of the spacecraft rotation to obtain a full 3D ion distribution within one spacecraft spin, cMISP on the Gateway, which is a 3-axis stabilized space station, requires the addition of electrostatic deflection plates at the instrument entrance to scan the FOV over a 360° ×120° solid angle (±60° with respect to the central entrance plane).

### 4.3.5  cMESP: Magnetospheric Electron Spectrometer

cMESP is an electron spectrometer that will determine the velocity distribution functions (VDF) of the solar wind electrons (pristine or reflected from lunar crustal magnetic field anomalies) and of the plasma sheet electrons, when the Gateway gets into the terrestrial magnetosphere.

A top-hat electrostatic analyzer instrument covering the ~5 eV to ~20 keV energy range is proposed. As for the cMISP instrument, the addition of electrostatic deflection plates at the instrument entrance to scan the FOV over a 360° ×120° solid angle is required.

The proposed cMESP instrument is based on the SWEA (Solar Wind Electron Analyzer) instrument, flying onboard the MAVEN spacecraft (Mitchell et al., 2016).

### 4.3.6  cENPD: Energetic Particles Detector

The cENPD instrument will detect and measure the fluxes of the energetic charged particles, ions and electrons: Solar Energetic Particles (SEPs), low-energy Galactic Cosmic Rays (GCRs) and terrestrial plasma sheet energetic particles. The instrument will also investigate the spectra of the secondary high energy ions, released from the lunar surface following its irradiation by GCRs and/or SEPs (albedo energetic particles, cf. section 2.1). It will cover the ~40 keV – ~100 MeV energy range for ions and ~20 keV – ~30 MeV for electrons. It will provide a $\Delta E/E \leq 10$ keV energy resolution and supply, for ions, a measure of the composition (protons to iron nuclei).

In order to cover both pristine and albedo energetic particles, it will consist of two identical detection heads, each with a 60° × 60° FOV: one pointing to the lunar zenith and the other pointing to the opposite direction (lunar nadir). Each detection head will be composed of a collimator and a 1 cm$^2$, 1 mm thick silicon detector. In front of the detector a filter wheel will allow to place either a thick foil, a pinhole or an obturator to allow the reconfiguration of the detection head to various scientific modes to measure the combined spectra of electrons and ions, to measure the electron spectrum, to protect the detector from sunlight or to avoid saturation of the detector.

The proposed cENPD instrument will benefit from the heritage of the IPD instrument, flown onboard the DEMETER satellite (Sauvaud et al., 2006), and of the IDEE instrument, developed for the TARANIS satellite (Lefeuvre et al., 2008).

### 4.3.7  cGCRD: Galactic Cosmic Rays Detector

The cGCRD instrument will measure the spectra and the composition of the Galactic Cosmic Rays and that of the Solar Energetic Particles, covering the 0.1 to ~5 GeV energy range. It will thus be complementary to the cENPD instrument, covering the higher energies.

The proposed cGCRD instrument is the Mini.PAN penetrating particle analyzer, which is an approved H2020-FETOPEN project that will build a demonstrator of the Penetrating particle ANalyzer (PAN) for deep space applications (Wu et al., 2019).

Mini.PAN is based on the particle detection principle of a magnetic spectrometer, with novel layout and detection concepts to optimize the measurement precision for both high flux and low flux particles. As above several hundred MeV/nuc standard methods for measuring particle energies (TOF, d$E$/d$x$, $\Delta E$-$E$) become less efficient, the use of magnetic spectrometry (the charged particle energy is derived from the degree of bending of its trajectory in the magnetic field) is used as the principal particle analysis method. In Mini.PAN the bending of the particle in the magnetic field is measured by precise silicon strip tracking detectors, while the elemental identity of the particle is determined by its charge and $Z$, which is measured with the d$E$/d$x$ method at multiple points. Mini.PAN is designed to precisely measure the momentum, the charge, the direction and the time of energetic particles between 100 MeV/nuc and a few GeV/nuc.

Mini.PAN offers much higher energy resolution (compared to integral measurements), especially in the > 100 MeV range, and is appropriate for precision energy and species measurements in the 100 MeV/nuc to low GeV/nuc range, which contains both albedo particles and the low energy part of the ambient GCR spectrum. This part is not well resolved by past solar wind observatories (e.g. ACE) or by massive GCR detectors in low Earth orbit (e.g. PAMELA). Mini.PAN is also a new type of miniaturized, advanced energetic particle detector that can be adapted and adjusted for deep space missions, where mass limitations exist.

As a bonus, the proposed concept for the cGCRD detector can also detect MeV ENAs (likely of heliospheric origin, cf. Mewaldt et al., 2009), because the detection method combines a strong magnet, the $\Delta E$-$E$ technique and particle tracking through successive, pixelated SSDs. Few-MeV hydrogen ENAs would give the characteristic $\Delta E$-$E$ signal on the SSD stack, but across a straight-line trajectory, since the magnet does not influence them, i.e. they can be separated from charged species (which also get detected) and from the very high energy GCRs (which are less detected but penetrate deeper).

### 4.3.8  cHENA: High-Energies ENA Imager

cHENA is a high-energies ENA (Energetic Neutral Atoms) imager, for detecting and imaging the ENAs produced by charge exchange interactions between the terrestrial plasma sheet or ring current energetic ions and the geocorona neutral hydrogen atoms. It will cover the ~10 – 500 keV energy range, and will be equipped with a collimator to both delimit the FOV (120° × 90° or narrower) and reject the charged particles. The transmitted ENAs then go through a TOF system and are detected by an MCP (64 × 64 pixels). Pointing the instrument optical axis towards the terrestrial inner magnetosphere requires mounting cHENA on an azimuthal (1-axis) articulation.

The proposed cHENA instrument is based on the MIMI-INCA ENA imager, flown onboard Cassini (Krimigis et al., 2004) and the HENA ENA imager, flown onboard the IMAGE mission (Mitchell et al., 2000).

### 4.3.9  cMENA: Medium-Energies ENA Imager

cMENA is a medium-energies ENA imager, for detecting and imaging the ENAs produced by charge exchange interactions between the terrestrial plasma sheet ions and the geocorona neutral hydrogen atoms. It will thus be complementary to the cHENA instrument, extending the coverage to lower energies (~1 keV – 100 keV). An additional objective for MENA is the detection and imaging of ENAs produced in the lunar environment, from the charge exchange interactions between the solar wind protons and the lunar exosphere. This requires flexibility in the instrument pointing (Earth or Moon pointing), which implies also mounting cMENA on its own azimuthal (1-axis) articulation.

The proposed instrument has a 90° × 10° instantaneous FOV and it uses, as cHENA, a collimator to both delimit the FOV and reject the charged particles. The collimator includes also a UV filter. The instrument provides a 5° × 10° angular resolution, i.e. one-dimensional images. It is based on the heritage of the wide-angle imaging neutral-atom spectrometer onboard the TWINS mission (McComas et al., 2009b) and of the SERENA-ELENA neutral atom imager onboard the MPO Mercury Planetary Orbiter of the BepiColombo mission (Orsini et al., 2021).

### 4.3.10 cLENA: Low-Energies ENA Imager

cLENA completes the suite of ENA imagers by covering the lowest energies (down to ~10 eV). These low-energy ENAs have two main sources: the charge exchange interactions of the solar wind protons with the lunar exosphere and the charge exchange interactions with the lunar surface. Moon pointing for its FOV is thus required.

The proposed instrument has a 15° × 15° field-of-view consisting of a single pixel and uses a conversion surface to ionize incoming ENAs and then feed them into an electrostatic wave system, which acts as a filter to pass only particles within the proper energy range. The particles then go through a TOF system. The instrument is capable of high-cadence observations of the solar wind - lunar surface interaction within the ~10 eV to ~3.3 keV energy range and with a ~50% $\Delta E / E$ energy resolution. It is based on the LNT instrument that has been designed for the Luna-Resurs-Orbiter (Luna 26) mission.

### 4.3.11 cUVIS: UV Imaging Spectrometer

cUVIS is a UV / EUV imaging spectrometer, sensitive to specific emission lines for observing the terrestrial exosphere (H: 121.6 nm, He: 58.4 nm, O: 130.4 nm, and N: 120.0 nm), the terrestrial plasmasphere (He$^+$: 30.4 nm, O$^+$: 83.6 nm) and the lunar exosphere (He: 58.4 nm, plus emission lines of other elements). It will thus cover the 30 – 130 nm wavelength range and will have a 0.1° × 7.5° FOV with a ~5 arcmin angular resolution. This resolution corresponds to about 0.15 $R_E$ (Earth radii) at the plasmasphere, as seen from the Moon.

Earth pointing requires, as for cHENA and for cMENA, mounting the instrument on its own azimuthal (1-axis) articulation. The articulation allows also pointing the instrument to the Moon, as a function of the scientific target of each observation session.

The proposed instrument is based on the heritage of the PHEBUS UV / EUV imaging spectrometer onboard the MPO Mercury Planetary Orbiter of the BepiColombo mission (Chassefière et al., 2010).

### 4.3.12 cLPEF: Langmuir Probe and E-field

cLPEF is a Langmuir probe instrument for providing ambient plasma diagnostics: a conductive probe, either biased or floating, is immersed into the plasma and the resulting electron / ion fluxes to the conducting surface provide electric current or voltage measurements with respect to the spacecraft. From these measurements the main plasma characteristics can be derived, including the plasma density, the electric field or the spacecraft floating potential. In order to provide unperturbed plasma measurements the probe has to be located well outside of the plasma sheath that forms around the spacecraft with a thickness proportional to the local Debye length.

The proposed cLPEF instrument will employ two spherical probes (~ 8 – 10 cm in diameter), each placed on the tip of a retractable boom. Since this requirement is identical to the one for the two main magnetometer sensors of the cMAGF instrument, and in order to optimize the resources and simplify the interfaces of the whole space plasma package, it is proposed to combine the sensors of these two instruments and house a tri-axial fluxgate magnetometer sensor within each of the two Langmuir spherical probes (cf. section 4.4 for the CAD figures). Such a combined Langmuir probe / magnetometer concept has been originally introduced as a part of an integrated plasma and dust package study conducted under the ESA Contract No. 4000103352/11/NL/AF in the framework of the proposed Lunar Lander mission, and it is the approach used on ESA's Comet Interceptor mission (Ratti et al., 2022).

An additional possibility is to mount occasionally a stand-alone Langmuir probe at the edge of the Gateway external robotic manipulator, so as to use this robotic arm in order to investigate the properties of the plasma sheath, forming around the different Gateway modules, at various locations.

Langmuir probes benefit from the heritage of instruments that have flown on several space missions, including the RPWS instrument onboard Cassini (Gurnett et al., 2004), ISL onboard the DEMETER satellite (Lebreton et al., 2006), DSLP onboard the PROBA-2 satellite, etc.

### 4.3.13  cWAVE: Waves Radio Instrument

cWAVE is an electromagnetic waves instrument for the study of terrestrial AKR (auroral kilometric radiation) emissions, occurring in the auroral region. It would then take advantage of the Moon occultation method, which was first implemented by the Radio Astronomy Explorer-2 mission (Kaiser and Alexander, 1976). An additional objective is the study of the radio emissions emitted by accelerated particles in the solar corona and the solar wind.

The proposed cWAVE instrument will measure the AC electric field (one component: fast electric waveform at 16.5 MHz, decimated electric waveform at 64.5 kHz) and the AC magnetic field (one component: magnetic waveform at 64.5 kHz). These three products will be delivered as waveform (event mode) and / or as averaged spectra (survey mode, with onboard FFT computation).

As for the cMAGF and the cLPEF instruments, the cWAVE sensor needs to be placed at the tip of a dedicated retractable boom.

Radio waves instruments benefit from the heritage of instruments that have flown on several space missions, including STAFF onboard Cluster (Cornilleau-Wehrlin et al., 2003), RPWS instrument onboard Cassini (Gurnett et al., 2004), etc.

### 4.3.14  Retractable booms

As indicated above, the mounting of the sensors of the combined cMAGF and cLPEF instruments and of the cWAVE instrument requires a total of three retractable booms, ~3 m each. These booms need indeed to be stowed at the beginning and at the end of the mission to allow the instruments to stay within the allocated envelope and to be transferred by the Airlock. We propose to use the compact deployable and retractable boom that has been developed by Oxford Space Systems: the Astrotube Boom. It can be deployed to up to 3 m and is TRL 9.

### 4.3.15  Instruments not included in the conceptual design study

The above-described model payload instruments (cf. also **Table 5**) cover satisfactorily the instrumentation requirements, as defined by the topical team (cf. **Table 4**). However, there are two instruments that were not included in this conceptual design study: the MeV ENA Imager and the Soft X-ray Imager.

The MeV ENA Imager was not included due to the absence in Europe, in our knowledge, of a developed instrument or protype, for ENAs at these very high energies. However, as described above, the proposed cGCRD instrument (Mini.PAN) will be able to detect few-MeV hydrogen ENAs, separating them from similar energy protons and providing 1-pixel images of this population.

For X-ray imaging, a soft X-ray imager with a wide field-of-view, using lobster-eye optics and a position-sensitive MCP detector operating at the $0.1 - 2$ keV X-ray bandpass has been considered (cMXRI instrument). This instrument is based on the DXL/STORM soft X-ray imager protype flown onboard a sounding rocket mission (Collier et al., 2015). However, the size of this instrument (78 cm length) appears to be incompatible with the Gateway interfaces for mounting external payloads. This suggests that the X-ray imager could not be accommodated as part of this instruments package. However, it is recommended to propose cMXRI as a payload for the Large European Lander for the Moon.

## 4.4 Instrument accommodation

The CAD model for the instrument conceptual design and their accommodation on the Gateway was established in cooperation with the CNES PASO (Plateau d'Architecture des Systèmes Orbitaux), with the help of its concurrent engineering facilities (CIC : Centre d'Ingénierie Concourante), and particularly by using the IDM-CIC (Integrated Design Model) and IDM-View tools (https://idm.virtual-it.fr/).

The instrument accommodation on the Gateway modules has to fulfil several requirements:

– In-situ measurement low-energy plasma instruments have to be placed on areas with low electrostatic charging (cf. section 4.2).
– Pointing requirements for instruments with a field-of-view (cf. **Table 5**).
– Unobstructed field-of-view for these instruments.
– cGCRD, which has a strong permanent magnet perturbing the low-energy plasma measurements, should not be placed close to these instruments.
– Instrument grouping, when possible, to form self-contained "instrument suites", with instruments mounted on a common platform, minimizing interfaces with the Gateway and using a single SORI (external Small ORU Interface) for attachment on the Gateway.

The instrument accommodation configuration we propose, and is compatible with the above requirements, has the instruments grouped on one main and one secondary platform. Each of these two platforms, of the order of 0.8 m × 0.8 m, is mounted externally on a SORI attachment and is double sided, i.e. has instruments mounted on both sides of the platform. This accommodation is of course notional and could be subject to modifications, depending on Gateway engineering and programmatic constraints.

Both platforms are on the Logistics Module and they are mounted on two diametrically opposite positions, on the +X side (Main Instrument Platform) and on the –X side (Secondary Instrument Platform) of it, cf. **Figure 16**.

The two sides of the Main Platform are shown in **Figure 17**. With its positioning on the +X side of the Logistics Module, the Main Platform provides an unobstructed view to the solar wind arrival direction (**Figures 13C** and **13D**) and takes advantage of a favorable electrostatic environment (**Figure 15**). It is thus well suited for mounting the solar wind instruments (cSWFC and cSWIS), shown in **Figure 18** and **Figure 19** respectively.

The Main Instrument Platform hosts also:

– The magnetospheric particle instruments cMESP and cMISP, shown in **Figure 18** and **Figure 19** respectively.
– The energetic particle detector cENPD, shown in **Figure 20**, which is mounted on the –Y edge of the platform. cENPD has two oppositely directed FOVs, one along the +Z axis and one along the –Z axis. In this way, during periapsis passes one of the FOVs looks in the zenith direction, to monitor the pristine energetic particles precipitating towards the Moon's surface, whereas the other looks in the nadir direction, to monitor the albedo energetic particles that are the result of the interaction of the precipitating energetic particles with the lunar regolith (cf. section 4.5). Moreover, the +Z / –Z orientation of the two detector heads allows avoiding direct sunlight entering the detectors (Sun is in the +X direction).

– The two "compact" remote sensing instruments cMENA and cLENA, shown in **Figure 18** and **Figure 19** respectively. cMENA uses a dedicated azimuthal (1-axis) articulation. The cLENA orientation gives access, during the periapsis passes, to the Moon surface and plasma environment.

– The two booms of the fluxgate magnetometer package (cMAGF), which as described in section 4.3.1 consists of three different types of units. The main type is two pieces of boom-mounted dual fluxgate magnetometers (one sensor at each of the two ~3 m boom tips and one at the boom root). These two retractable booms are mounted on the Main Instrument Platform –Z side (**Figure 18**). The boom tip mounted cMAGF sensors are integrated together with Langmuir probes (cLPEF instrument). These units will provide the main measurements. In order to monitor the perturbations from the station close to the source in more detail, several (~5+) single magnetometer sensors will be also mounted on various places around the station (not shown).

– The cWAVE instrument, also mounted on a retractable boom, which is on the Main Instrument Platform +Z side (**Figure 19**).

The two sides of the Secondary Platform are shown in **Figure 21**. This platform, mounted on the –X side of the Logistics Module, is permanently in the shadow. In this way there is no direct sunlight that could interfere with the measurements of the two instruments mounted on it. On each of its two sides there is a remote sensing instrument: cUVIS on the one side and cHENA on the other side of the Secondary Platform. Each of these two instruments in mounted on a dedicated azimuthal (1-axis) articulation.

The cGCRD instrument, due to the containment of a strong magnet (0.4 Tesla) that would deviate charged particles to be measured by the other instruments if in close vicinity with them, is not mounted on any of the two instrument platforms. It is instead mounted as a "standalone" instrument on the SORI attachment of the +Z side of the Logistics Module (**Figure 22**). Its FOV, looking radially out, near periapsis gives access to the albedo energetic particles that are the result of the interaction of the precipitating galactic cosmic ray particles with the lunar regolith. During the remaining part of the orbit (most of the time) it points to the open sky.

## 4.5 Instruments fields-of-view simulation

The appropriate orientation of the fields-of-view (FOVs) of the remote sensing and of the high-energy particle instruments, as projected on the sky and on the celestial objects, was analyzed by simulating the evolution of the FOVs along the Gateway orbit. This simulation was performed in cooperation with the CNES PASO and by using the VTS software tool (https://logiciels.cnes.fr/en/content/vts).

The FOVs of the two oppositely directed sensor heads of the cENPD instrument, near a periapsis pass, are shown in **Figure 23**. As shown in this figure, one of the two sensor heads is oriented towards the local zenith, and has an unobstructed view to the pristine energetic particles flux (purple FOV), whereas the other is oriented towards the nadir and its FOV is dominated by the albedo energetic particles from the Moon (yellow FOV). Both populations (pristine and albedo high-energy particles) are thus covered by the cENPD instrument detection capabilities.

For the cGCRD instrument, which is a single sensor head GCR detector, the FOV near a periapsis pass is shown in **Figure 24**, left panel (light blue FOV). As shown, near periapsis it is dominated by the albedo GCR particles from the Moon. However, during most of the remaining orbit (right panel) it has an unobstructed view to the open sky and gives then access to the pristine GCR environment.

The field-of-regard (FOR) of the cMENA instrument, i.e. the total accessible FOV taking into account the rotation of the 1-axis articulation on which the instrument is mounted, at a given point of the orbit, is shown in **Figure 25A**. The azimuthal rotation mechanism gives to the instrument access to a very large "ribbon" of the sky, which includes the Earth environment and the Moon environment. The pointing of the instrument to any of these two principal targets, using the flexibility provided by the 1-axis articulation, can then be programmed as a function of the scientific target of each observation session.

In **Figure 25B** is the FOV of the cLENA instrument, close to periapsis, as projected on the sky (no articulation for this instrument). As shown in this figure, the way the instrument is mounted on the Gateway gives access, during the periapsis passes, to the Moon surface and to its exosphere and plasma environment.

The FOR of the cUVIS instrument is shown in **Figure 25C**. As shown in this figure, the dedicated articulation allows also for this instrument to point to targets as the Earth space environment (plasmasphere, exosphere), the Moon space environment (exosphere), or targets in the open sky. The narrow width of the instantaneous FOV of this instrument (0.1°), in combination with the articulation, allows also performing altitude profile scans of the lunar exosphere.

## 5    Conclusion

The Moon is a unique location to study the deep space plasma environment. The Lunar Orbital Platform - Gateway, that will be assembled and operated in the vicinity of the Moon starting from the mid-2020s, is a crewed station that offers new opportunities for fundamental and applied scientific research in the field of space plasma physics. These have multi-disciplinary dimensions, and they include:

–   Studying the lunar space environment and its interaction with the solar wind and the terrestrial magnetotail plasma.
–   Terrestrial space weather: monitoring, through remote sensing techniques, the response of the terrestrial magnetosphere and exosphere to solar activity events.
–   Planetary space weather: monitoring, through in-situ measurements and through remote sensing, the response of the lunar space environment to solar activity events.
–   Radiation physics: characterizing the lunar high-energy particles environment, including energy and mass spectrometry of these populations and their variability, particularly in view of the Artemis human missions to the Moon and the associated radiation risks.
–   Studying the heavy ion escape from the terrestrial ionosphere, through in-situ measurements of the downtail streaming ions, and the role of this escape in the long-term evolution of the composition of the terrestrial atmosphere (and its habitability).
–   Studying the lunar regolith - bounded exosphere - interplanetary space environment as a complex interacting multi-scale system, and as an archetype of the interaction of an unmagnetized planetary body with the solar wind.
–   Studying the mini-magnetospheres that form above the "swirls" on the Moon, and which constitute probably the smallest magnetospheres in our solar system.

- Understanding the surface electric fields that develop on the Moon as a part of a complex and interacting plasma environment, and their role in electrostatic lunar dust levitation.
- Planetology: understanding the composition of the lunar regolith, and its hydration, through the spectrometry of the albedo energetic particles.

In preparation of the scientific payload of the Lunar Orbiter Platform - Gateway we first formed a topical team, under the auspices of ESA, to prepare and to support the definition of payload studies in the field of space plasma physics. This allowed to identify the scientific objectives that can be addressed from onboard the Lunar Orbital Platform - Gateway, the physical parameters needed to be measured in order to address these objectives, and the corresponding instrumentation required to perform these in-situ measurements and remote-sensing observations.

We then undertook for ESA a conceptual design study for a "Space Plasma Physics Payload Package onboard the Gateway" (SP4GATEWAY), addressing the objectives identified by the topical team while remaining compatible with the technical requirements. This conceptual design has considered, as baseline, a typical "Gateway Phase 2" configuration.

As a first part of this conceptual design study, we simulated the interaction between the Gateway and its plasma environment, for the case where the Gateway is in the solar wind and also for the case where the Gateway is in the terrestrial magnetotail. This allowed to identify the Gateway modules on which the perturbation of the natural plasma environment by the Gateway will be minimal, and are thus best-suited for placing there the in-situ measurement plasma instruments.

We then defined a model payload consisting of a suite of instruments, for in-situ measurements and for remote-sensing observations, corresponding to the requirements. These measurement instruments are largely based, either on existing flight-proven instruments, adapted here for the lunar plasma environment, or on tested and validated laboratory prototypes. The main characteristics of these instruments have been defined and CAD conceptual instrument designs elaborated. The instruments' measurement characteristics will however have to be refined during a follow-on Phase A study.

The next step was the study for accommodating this model payload on the Gateway modules, taking into account the various constraints, and in particular the surface and volume charging of the various Gateway modules, their exposure to the ambient plasma and the pointing and field-of-view requirements of the instruments. This resulted in an integrated CAD design, including the Gateway and the instruments, which were grouped into two platforms mounted on two sides of the Logistics Module.

The fields-of-view of the remote sensing instruments and of the high-energy particle instruments, as projected on the sky and on the celestial objects, were then analyzed by simulating their evolution along the Gateway orbit. This allowed to verify the appropriate orientation of the fields-of-view and the coverage of the observational scientific targets.

Following this conceptual design study for a Space Plasma Physics Payload Package onboard the Gateway, it results that the Gateway is very well-suited for space plasma physics research and it allows to address a series of relevant scientific objectives.

# 6    Tables

**TABLE 1 | Physical parameter / observable to be monitored from onboard the Gateway**

| | |
|---|---|
| In-situ measurements | Solar Wind (particles + fields) |
| | Earth's foreshock |
| | SEPs, GCRs (pristine + secondary from Moon, at various directions) |
| | Energetic electrons |
| | Magnetotail + magnetosheath plasma  (particles + fields) |
| | Outflowing terrestrial ions  (ion spectrometry) |
| | Lunar pickup ions  (ion spectrometry) |
| | Gateway-induced plasma and fields environment |
| | Lunar Wake |
| Imaging | MeV ENAs: produced from SEPs |
| | ENAs:  Terrestrial Ring Current and Plasma Sheet |
| | Low-energy ENAs (from Solar Wind and Moon) |
| | SWCX X-rays:  Magnetosheath/pause + cusp +  planetary targets of opportunity |
| | UV / EUV:  Terrestrial Plasmasphere |
| | UV / EUV:  Geocorona,  Lunar Exosphere,  Solar EUV radiometry |
| | Auroral imaging,  Planetary imaging |
| | Heliosphere imaging |
| | Lunar surface micrometeorite impacts |
| Active experiments | Gas release and ionization |

| **TABLE 2** | Physical parameter / observable to be monitored from low lunar orbits | |
|---|---|
| In-situ measurements | Crustal Magnetic Anomalies  (Plasma + magnetic field + ENAs + electron reflectometry) |
| | Solar wind ions neutralization |
| | Lunar Exosphere / Ionosphere  (in-situ measurements) |
| | Dusty plasmas |
| Imaging | Lunar Exosphere / Ionosphere  (imaging) |

| **TABLE 3** | Physical parameter / observable to be monitored from the lunar surface | |
|---|---|
| In-situ measurements | Energetic ion implantation / reflection |
| | Lunar surface electrostatic charging + dust |
| | Crustal Magnetic Anomalies |
| | Magnetosphere radio emissions |
| | Lunar exosphere |
| Imaging | SWCX X-rays:  Magnetosheath/pause + cusp + planetary targets of opportunity |

**TABLE 4 |** Science objectives and corresponding measurement and instrumentation requirements (from onboard the Gateway)

| Science Objective | Measurement Requirement | In-situ Measurements Instrument | Remote Sensing Instrument |
|---|---|---|---|
| Monitor solar wind as a driver for the dynamics of terrestrial magnetosphere, terrestrial and lunar exospheres, lunar surface sputtering and charging | Solar wind density and transport velocity $1 - 10^2$ cm$^{-3}$, $0.1 - 40$ keV ions $200 - 1000$ km/s, $\Delta E/E < 17\%$ | Faraday Cup Electrostatic Analyzer | - |
| | IMF: 100 nT instrument range 1 nT / 0.1 nT absolute / relative resolution | Magnetometer | - |
| Monitor and characterize SEPs and GCRs for radiation environment and as lunar surface sputtering sources | 40 keV – 100 MeV ions (SEPs) up to ~5 GeV (GCRs) 50 MeV / nucleon for composition ~40 keV – ~30 MeV electrons | Energetic particle detectors | MeV ENA Imager |
| Monitor and characterize the response of the terrestrial magnetosphere to the solar wind with a wide coverage of geospace | Detect and image solar wind charge exchange X-rays $0.2 - 2.0$ keV FOV $10° \times 10°$ angular resolution: 0.3 R$_E$ from the Moon | - | Soft X-ray Imager |
| | Detect and image terrestrial magnetosphere ENAs $\sim 1 - 300$ keV, FOV $\sim 20° \times 20°$ | | ENA Imager |
| Monitor solar wind interaction with the lunar exosphere, regolith and magnetic anomalies | Detect and image low-energy ENAs: $0.1 - 10$ keV, 30 % $\Delta E/E$, FOV $\sim 20° \times 20°$, ~5° resolution Strong UV suppression: $10^{-8}$ | - | LENA imager |
| Reveal the solar wind ion dynamics in the vicinity of the lunar magnetic anomalies | Detect and image low-energy ENAs: $0.01 - 3$ keV, 30 % $\Delta E/E$, FOV $\sim 5° \times 120°$, ~5° resolution | - | LENA imager |
| Monitor the terrestrial and lunar exospheres, plasmasphere | Detect and image EUV emissions 30.4, 83.6, 121.6 and 130.4 nm ~5 arcmin resolution | Ion mass spectrometer (lunar pickup ions) | UV / EUV spectro-imager |
| Monitor ambient plasma in different environments (solar wind / magnetosheath / terrestrial magnetotail / lunar wake) | Plasma density and temperature ~0.01 – 40 keV, $10^{-3} - 10^2$ cm$^{-3}$ Ion composition: $m/\Delta m > 15$ | Langmuir probe Ion mass spectrometer Electron spectrometer | - |
| | Magnetic field: 1000 nT range 1 nT / 0.1 nT absolute/relative resolution | Magnetometer | - |
| Monitor magnetospheric and planetary radio emissions | | | Radio instrument |

**TABLE 5 | SP4Gateway model payload instruments**

| Instrument Acronym | Instrument | Mass (kg) | Power (W) | FOV | FOV pointing |
|---|---|---|---|---|---|
| cMAGF | Magnetometer(s) | 3.5 | 4.3 | N/A | N/A |
| cSWIS | Solar Wind Ion Spectrometer | 5 | 7 | 96° × 48° | Sun |
| cSWFC | Solar Wind Faraday Cup | 5 | 4 | 45° × 45° (×6) | Sun |
| cMISP | Magnetospheric Ion Spectrometer | 7 | 8 | 360° × 120° | N/A |
| cMESP | Magnetospheric Electron Spectrometer | 3 | 6 | 360° × 120° | N/A |
| cENPD | Energetic Particles Detector | 3 | 6 | 60° × 60° (×2) | Moon / Sky |
| cGCRD | Galactic Cosmic Rays Detector | 10 | 20 | 71° × 71° | Moon / Sky |
| cHENA | High-Energies ENA Imager | 15 | 12 | 120° × 90°, articulation | Earth |
| cMENA | Medium-Energies ENA Imager | 5 | 15 | 90° × 10°, articulation | Earth / Moon |
| cLENA | Low-Energies ENA Imager | 4 | 10 | 15° × 15° | Moon |
| cUVIS | UV Imaging Spectrometer | 10 | 15 | 0.1° × 7.5°, articulation | Earth / Moon |
| cLPEF | Langmuir Probe and E-field | 1.2 | 5 | N/A | N/A |
| cWAVE | Waves Radio Instrument | 5.6 | 8.6 | N/A | N/A |

*Mass and Power: nominal values, without margins, booms and articulation mechanisms included in these values; FOV: field-of-view; N/A: not applicable*

# 7 Conflict of Interest

The authors declare that the research was conducted in the absence of any commercial or financial relationships that could be construed as a potential conflict of interest.

# 8 Author Contributions

ID was the coordinator of the topical team "Space Plasma Physics Science Opportunities for the Lunar Orbital Platform - Gateway", the coordinator of the SP4GATEWAY project, and is the main author of this manuscript. MGGT was the ESA support scientist for the topical team and for the SP4GATEWAY project. JDK, YF, RAB, GBR, JYF, DG, BG, HL, FL, AM, RN, and ER were members of the topical team and of the SP4GATEWAY project team. PD was the lead engineer for the SP4GATEWAY project. EDA, JE, ME, PG, DH, LP and ŠŠ were members of the SP4GATEWAY project team. AL managed the CAD design. JF, AT, SLGH and JCMV performed the Gateway - plasma environment simulations. JC was the ESA HRE (Human and Robotic Exploration) correspondent. JW was the ESA manager for the SP4GATEWAY project.

# 9 Funding

The topical team "Space Plasma Physics Science Opportunities for the Lunar Orbital Platform - Gateway" was supported by ESA through contract No. 4000128802 /19/NL/PG/pt. The SP4GATEWAY project was funded by ESA through contract No. 4000128461/19/NL/FC. Activities at IRAP were also supported by CNES through order 4500065232.

# 10 Acknowledgments

We acknowledge the support of the CNES PASO for the SP4GATEWAYconceptual design study.

# 11 References

Allegrini, F., Dayeh, M. A., Desai, M. I., et al. (2013). Lunar energetic neutral atom (ENA) spectra measured by the interstellar boundary explorer (IBEX). *Planetary and Space Sci.*, 85, doi: 10.1016/j.pss.2013.06.014

André, M., Toledo-Redondo, S., and Yau, A. W. (2021). Cold Ionospheric Ions in the Magnetosphere. *Magnetospheres in the Solar System*, Geophysical Monograph 259, Chapter 15, doi: 10.1002/9781119815624.ch15

Angelopoulos, V. (2011). The ARTEMIS mission. *Space Sci. Rev.*, doi: 10.1007/s11214-010-9687-2

Artemis III Science Definition Team Report (2020), NASA report NASA/SP-20205009602

Auster, H. U., Glassmeier, K. H., Magnes, W., et al. (2008). The THEMIS Fluxgate Magnetometer. *Space Sci. Rev.*, 141, doi: 10.1007/s11214-008-9365-9

Balogh, A., Carr, C. M., Acuña, M. H., Dunlop, M. W., Beek, T. J., Brown, P., Fornacon, K.-H., Georgescu, E., Glassmeier, K.-H., Harris, J., Musmann, G., Oddy, T., and Schwingenschuh, K. (2001). The Cluster Magnetic Field Investigation: overview of in-flight performance and initial results. *Ann. Geophys.*, 19, 1207–1217, doi: 10.5194/angeo-19-1207-2001

Bamford, R. A., Alves, E. P., Cruz, F., Kellett, B. J., Fonseca, R. A., Silva, L. O., Trines, R. M. G. M., Halekas, J. S., Kramer, G., Harnett, E., Cairns, R. A.; and Bingham, R. (2016). 3D PIC

Simulations of Collisionless Shocks at Lunar Magnetic Anomalies and Their Role in Forming Lunar Swirls. *Astroph. J.*, doi: 10.3847/0004-637X/830/2/146

Benna, M., Mahaffy, P. R., Halekas, J. S., Elphic, R. C., and Delory, G. T. (2015). Variability of helium, neon, and argon in the lunar exosphere as observed by the LADEE NMS instrument. *Geophys. Res. Lett.*, doi: 10.1002/2015GL064120

Bosqued, J. M., Lormant, N., Rème, H., d'Uston, C., Lin, R. P., Anderson, K. A., Carlson, C. W., Ergun, R. E., Larson, D., McFadden, J., McCarthy, M. P., Parks, G. K., Sanderson, T. R., Wenzel, K.-P. (1996). Moon-solar wind interactions: First results from the WIND/3DP Experiment. *Geophys. Res. Lett.*, doi: 10.1029/96GL00303.

Brandt, P. C:son, Roelof, E. C., Ohtani, S., Mitchell, D. G., and Anderson, B. (2004). IMAGE/HENA: pressure and current distributions during the 1 October 2002 storm. *Adv. Space Res.*, doi: 10.1016/S0273-1177(03)00633-1

Branduardi-Raymont, et al. (2012). AXIOM: advanced X-ray imaging of the magnetosphere. *Exp. Astron.*, doi: 10.1007/s10686-011-9239-0

Branduardi-Raymont, G., Berthomier, M., Bogdanova, Y. V. et al. (2021). Exploring solar-terrestrial interactions via multiple imaging observers. *Exp. Astron.*, doi: 10.1007/s10686-021-09784-y

Burinskaya, T. M. (2015). Non-monotonic potentials above the day-side lunar surface exposed to the solar radiation. *Planet. Space Sci.*, doi: 10.1016/j.pss.2015.03.004

Cao, J., Ma, Y., Parks, G., Rème, H., Dandouras, I., and Zhang, T. (2013). Kinetic analysis of the energy transport of bursty bulk flows in the plasma sheet. *J. Geophys. Res.*, doi: 10.1029/2012JA018351

Cara, A., Lavraud, B., Fedorov, A., De Keyser, J., DeMarco, R., Marcucci, M. F., Valentini, F., Servidio, S., and Bruno, R. (2017). Electrostatic analyzer design for solar wind proton measurements with high temporal, energy, and angular resolutions. *J. Geophys. Res.*, doi: 10.1002/2016JA023269

Carpenter, J., Speak, C., Chouker, A., Talevi, M., Nakamura, R., Santangelo, A., Crawford, I., Cullen, D., Bussey, B., and Grenouilleau, J. (2018). Research Opportunities on the Deep Space Gateway: Findings from the Workshop and Call For Ideas, ESA-HSO-K-RP-0284

Chassefière, E., Maria, J.-L., Goutail, J.-P., Quémerais, E., Leblanc, F., et al. (2010). PHEBUS: A double ultraviolet spectrometer to observe Mercury's exosphere. *Planetary and Space Sci.*, doi: 10.1016/j.pss.2008.05.018

Chin, G., Brylow, S., Foote, M., Garvin, J., Kasper, J., Keller, J., Litvak, M., Mitrofanov, I., Paige, D., Raney, K., Robinson, M., Sanin, A., Smith, D., Spence H., Spudis, P., Stern, S. A., and Zuber, M. (2007). Lunar Reconnaissance Orbiter overview: The instrument suite and mission. *Space Sci. Rev.*, 129, 391–419, doi: 10.1007/s11214-007-9153-y

Christon, S. P., Gloeckler, G., Williams, D. J., Mukai, T., McEntire, R. W., Jacquey, C., Angelopoulos, V., Lui, A. T. Y., Kokubun, S., Fairfield, D. H., Hirahara, M., and Yamamoto, T. (1994). Energetic atomic and molecular ions of ionospheric origin observed in distant magnetotail flow-reversal events. *Geophys. Res. Lett.*, 21(25), 3023–3026

Christon, S. P., Hamilton, D. C., Mitchell, D. G., Plane, J. M. C., and Nylund, S. R. (2020). Suprathermal magnetospheric atomic and molecular heavy ions at and near Earth, Jupiter, and Saturn: Observations and identification *J. Geophys. Res.*, 125, doi: 10.1029/2019JA027271

Collier, M. R., et al. (2015). First flight in space of a wide-field-of-view soft x-ray imager using lobster-eye optics: Instrument description and initial flight results. *Rev. Sci. Instrum.*, doi: 10.1063/1.492725

Constantinescu, O. D., Auster, H.-U., Delva, M., Hillenmaier, O., Magnes, W., and Plaschke, F. (2020). Maximum-variance gradiometer technique for removal of spacecraft-generated disturbances from magnetic field data. *Geosci. Instrum. Method. Data Syst.*, doi: 10.5194/gi-9-451-2020

Cornilleau-Wehrlin, N., Chanteur, G., Perraut, S., Rezeau, L., et al. (2003). First results obtained by the Cluster STAFF experiment. *Ann. Geophys.,* doi: 10.5194/angeo-21-437-2003

Crider, D. H., and Vondrak, R. R. (2003). Space weathering effects on lunar cold trap deposits. *J. Geophys. Res.*, doi: 10.1029/2002JE002030

C:son Brandt, P., Mitchell, D. G., Ebihara, Y., Sandel, B. R., Roelof, E. C., Burch, J. L., and Demajistre, R. (2002). Global IMAGE/HENA observations of the ring current: Examples of rapid response to IMF and ring current-plasmasphere interaction. *J. Geophys. Res.*, doi: 10.1029/2001JA000084

Dandouras, I., Yamauchi, M., De Keyser, J., et al. (2018). ESCAPE: a mission proposal for ESA-M5 to systematically study Exosphere and atmospheric escape using European, Japanese, and US instruments. *Proc. ISAS Symposium*, Japan 2018 https://repository.exst.jaxa.jp/dspace/handle/a-is/876320

Dandouras, I., Blanc, M., Fossati, L. et al. (2020a). Future Missions Related to the Determination of the Elemental and Isotopic Composition of Earth, Moon and the Terrestrial Planets. *Space Sci. Rev.,* 216, doi: 10.1007/s11214-020-00736-0

Dandouras, I., Bamford, R. A., Branduardi-Raymont, G., Chaufray, J.-Y., Constantinescu, D., De Keyser, J., Futaana, Y., Lammer, H., Milillo, A., Nakamura, R., Roussos, E., Grison, B., and Taylor, M. G. G. T. (2020b). Report of the ESA Topical Team: "Space Plasma Physics Science Opportunities for the Lunar Orbital Platform-Gateway". *ESA report*

Dandouras, I. (2021). Ion outflow and escape in the terrestrial magnetosphere: Cluster advances. *J. Geophys. Res.*, doi: 10.1029/2021JA029753

Darrouzet, F., Gallagher, D. L., André, N. et al. (2008). Plasmaspheric Density Structures and Dynamics: Properties Observed by the CLUSTER and IMAGE Missions. *Space Sci. Rev.*, doi: 10.1007/s11214-008-9438-9

Deca, J., Divin, A., Lembège, B., Horányi, M., Markidis, S., and Lapenta, G. (2015). General mechanism and dynamics of the solar wind interaction with lunar magnetic anomalies from 3-D particle-in-cell simulations, *J. Geophys. Res.*, doi :10.1002/2015JA021070

De Keyser, J., Lavraud, B., Přech, L., Neefs, E., Berkenbosch, S., Beeckman, B., Fedorov, A., Marcucci, M. F., De Marco, R., and Brienza, D. (2018). Beam tracking strategies for fast acquisition of solar wind velocity distribution functions with high energy and angular resolutions. *Ann. Geophys*, 36, 1285–1302, doi: 10.5194/angeo-36-1285-2018

De Keyser, J., et al., (2020). Space Physics Payload for the Deep Space Gateway (SP4GATEWAY): Requirements Inventory. *ESA report*, SP4GATEWAY-REQ-i01r05

Devoto, P., Médale, J.-L., and Sauvaud, J.-A. (2008). Secondary electron emission from distributed ion scattering off surfaces for space instrumentation. *Rev. Sci. Instr.*, 79, 046111, doi: 10.1063/1.2912821

Devoto, P., and Dandouras, I. (2020). SP4Gateway Conceptual Design Report. *ESA report*, SP4G-IRAP-TN-001

Dougherty, M. K., Kellock, S., Southwood, D. J. et al. (2004). The Cassini Magnetic Field Investigation. *Space Sci. Rev.*, 114, doi: 10.1007/s11214-004-1432-2

Du, A. M., Nakamura, R., Zhang, T. L., Panov, E. V., Baumjohann, W., Luo, H., Xu, W. Y., Lu, Q. M., Volwerk, M., Retinò, A., Zieger, B., Angelopoulos, V., Glassmeier, K.-H., McFadden, J. P., and Larson, D. (2011). Fast tailward flows in the plasma sheet boundary layer during a substorm on 9 March 2008: THEMIS observations. *J. Geophys. Res.*, doi: 10.1029/2010JA015969

Elphic, R. C., Delory, G. T., Hine, Butler P., Mahaffy, P. R., Horanyi, M., Colaprete, A., Benna, M., and Noble, S. K. (2014). The Lunar Atmosphere and Dust Environment Explorer Mission. *Space Sci. Rev.*, doi: 10.1007/s11214-014-0113-z

Farrell, W. M., Stubbs, T. J., Vondrak, R. R., Delory, G. T., and Halekas, J. S. (2007). Complex electric fields near the lunar terminator: The near-surface wake and accelerated dust. *Geophys. Res. Lett.*, doi: 10.1029/2007GL029312

Farrell, W. M., Hurley, D. M., Esposito, V. J., McLain, J. L., and Zimmerman, M. I. (2017). The statistical mechanics of solar wind hydroxylation at the Moon, within lunar magnetic anomalies, and at Phobos. *J. Geophys. Res.*, doi: 10.1002/2016JE005168

Futaana, Y., Barabash, S., Holmström, M., and Bhardwaj, A. (2006). Low energy neutral atoms imaging of the Moon. *Planetary and Space Sci.*, 54, doi: 10.1016/j.pss.2005.10.010

Futaana, Y., Nakano, S., Wieser, M., and Barabash, S. (2008). Energetic neutral atom occultation: New remote sensing technique to study the lunar exosphere. *J. Geophys. Res.*, doi: 10.1029/2008JA013356

Futaana, Y., et al. (2012). Empirical energy spectra of neutralized solar wind protons from the lunar regolith. *J. Geophys. Res.*, doi: 10.1029/2011JE004019

Futaana, Y., Barabash, S., Wieser, M., Lue, C., Wurz, P., Vorburger, A., Bhardwaj, A., and Asamura, K. (2013). Remote energetic neutral atom imaging of electric potential over a lunar magnetic anomaly. *Geophys. Res. Lett.*, doi: 10.1002/grl.50135

Futaana, Y., et al. (2018). SELMA mission: How do airless bodies interact with space environment? The Moon as an accessible laboratory. *Planet. Space Sci.*, doi: 10.1016/j.pss.2017.11.002

Futaana, Y., et al. (2020). SP4Gateway Programmatic Assessment. *ESA report*, SP4G_IRF-TN-001

Gebauer, S., Grenfell, J. L., Lammer, H., Paul de Vera, J.-P., Sproß, L., Airpetian, V. S., Sinnhuber, M., and Rauer, H. (2020). Atmospheric nitrogen when life evolved on Earth. *Astrobiology*, 20, 1413–1426, doi: 10.1089/ast.2019.2212

Geiss, J., Bühler, F., Cerutti, H., Eberhardt, P., Filleux, Ch., Meister, J., and Signer, P. (2004). The Apollo SWC Experiment: Results, Conclusions, Consequences. *Space Sci. Rev.*, doi: 10.1023/B:SPAC.0000023409.54469.40

Glassmeier, K.-H., Auster, H.-U., Heyner, D., Okrafka, K., Carr, C., et al., (2010). The fluxgate magnetometer of the BepiColombo Mercury Planetary Orbiter. *Planetary and Space Sci.*, doi: 10.1016/j.pss.2008.06.018

Glotch, T., Bandfield, J., Lucey, P. et al. (2015). Formation of lunar swirls by magnetic field standoff of the solar wind. *Nature Commun.*, doi: 10.1038/ncomms7189

Goldstein, J., Valek, P. W., McComas, D. J., and Redfern, J. (2022). Average ring current response to solar wind drivers: Statistical analysis of 61 days of ENA images. *J. Geophys. Res.*, doi: 10.1029/2021JA029938

Grava, C., Chaufray, J.-Y., Retherford, K. D., Gladstone, G .R., Greathouse, T.K., Hurley, D. M., Hodges, R .R., Bayless, A. J., Cook, J. C., and Stern, S. A. (2015). Lunar exospheric argon modeling. *Icarus*, doi: 10.1016/j.icarus.2014.09.029

Grava, C., Retherford, K. D., Hurley, D. M., Feldman, P. D., Gladstone, G. R., Greathouse, T. K., Cook, J. C., Stern, S. A., Pryor, W. R., Halekas, J. S., and Kaufmann, D. E. (2016). Lunar exospheric helium observations of LRO/LAMP coordinated with ARTEMIS. *Icarus*, doi: 10.1016/j.icarus.2015.10.033

Grava, C., Killen, R. M., Benna, M. et al. (2021). Volatiles and Refractories in Surface-Bounded Exospheres in the Inner Solar System. *Space Sci. Rev.*, doi: 10.1007/s11214-021-00833-8

Grigorenko, E. E., Runov, A., Angelopoulos, V., and Zelenyi, L. M. (2019). Particle beams in the vicinity of magnetic separatrix according to near-lunar ARTEMIS observations. *J. Geophys. Res.*, 124, doi: 10.1029/2018JA026160

Grün, E., Horanyi, M., and Sternovsky, M. (2011). The lunar dust environment. *Planet. Space Sci.*, doi: 10.1016/j.pss.2011.04.005

Gurnett, D. A., Kurth, W. S., Kirchner, D. L., Hospodarsky, G. B., et al. (2004). The Cassini Radio and Plasma Wave Science Investigation, *Space Sci. Rev.*, 114, doi: 10.1007/s11214-004-1434-0

Halekas, J. S., Delory, G. T., Farrell, W. M., Angelopoulos, V., McFadden, J. P., Bonnell, J. W., Fillingim, M. O., and Plaschke, F. (2011). First remote measurements of lunar surface charging from ARTEMIS: Evidence for nonmonotonic sheath potentials above the dayside surface. *J. Geophys. Res.*, doi: 10.1029/2011JA016542

Halekas, J. S., Benna, M., Mahaffy, P. R., Elphic, R. C., Poppe, A. R., and Delory, G. T. (2015). Detections of lunar exospheric ions by the LADEE neutral mass spectrometer. *Geophys. Res. Lett.*, doi: 10.1002/2015GL064746

Hapke, B. (2001). Space weathering from Mercury to the asteroid belt. *J. Geophys. Res.*, 106, 10039–10073, doi: 10.1029/2000JE001338

Harada, Y., Poppe, A. R., Halekas, J. S., Chamberlin, P. C., and McFadden, J. P. (2017). Photoemission and electrostatic potentials on the dayside lunar surface in the terrestrial magnetotail lobes. *Geophys. Res. Lett.*, doi: 10.1002/2017GL073419

Harnett, E. M., Cash, M., and Winglee, R. M. (2013). Substorm and storm time ionospheric particle flux at the Moon while in the terrestrial magnetosphere. *Icarus*, doi: 10.1016/j.icarus.2013.02.022

Hartle, R. E., and Killen, R. (2006). Measuring pickup ions to characterize the surfaces and exospheres of planetary bodies: Applications to the Moon. *Geophys. Res. Lett.*, 33, doi: 10.1029/2005GL024520

He, F., Zhang, X.-X., Chen, B., Fok, M.-C., and Nakano, S. (2016). Determination of the Earth's plasmapause location from the CE-3 EUVC images. *J. Geophys. Res.*, doi: 10.1002/2015JA021863

Hemingway, D. J., Garrick-Bethell, I., and Kreslavsky, M. A. (2015). Latitudinal variation in spectral properties of the lunar maria and implications for space weathering. *Icarus*, doi: 10.1016/j.icarus.2015.08.004

Hemingway, D. J., and Tikoo, S. M. (2018). Lunar swirl morphology constrains the geometry, magnetization, and origins of lunar magnetic anomalies. *J. Geophys. Res.*, doi: 10.1029/2018JE005604

Hess, S. L. G., Sarrailh, P., Matéo-Vélez , J.-C., Jeanty-Ruard, B., Cipriani, F., Forest, J., Hilgers, A., Honary, F., Thiébault, B., Marple, S. R., and Rodgers, D. (2015). New SPIS Capabilities to Simulate Dust Electrostatic Charging, Transport, and Contamination of Lunar Probes. *IEEE Transactions on Plasma Science*, 43 (9), 2799–2807, doi: 10.1109/TPS.2015.2446199

Hess, S. L. G., Sarrailh, P., Villemant, M., and Trouche, A. (2020). Modélisation de l'environnement électrostatique de la Lunar Orbital Platform-Gateway. *ONERA - Artenum Report*, SPACESUITE-2020-IRAP-RF

Hodges, R. R., Jr. (2016). Methane in the lunar exosphere: Implications for solar wind carbon escape. *Geophys. Res. Lett.,* doi: 10.1002/2016GL068994

Honniball, C.I., Lucey, P.G., Li, S. et al. (2021). Molecular water detected on the sunlit Moon by SOFIA. *Nature Astron.*, doi: 10.1038/s41550-020-01222-x

Horányi, M., Szalay, J., Kempf, S. et al. (2015). A permanent, asymmetric dust cloud around the Moon. *Nature*, doi: 10.1038/nature14479

Hurley, D. H., Cook, J. S., Benna, M., Halekas, J. S., Feldman, P. D., Retherford, K. D., Hodges, R. R., Grava, C., Mahaffy, P., Gladstone, G. R., Greathouse T., Kaufmann, D. E., Elphic, R. C., and Stern, S. A. (2016). Understanding the temporal and spatial variability of the lunar helium atmosphere using simultaneous observations from LRO, LADEE and ARTEMIS. *Icarus*, 273, doi: 10.1016/j.icarus.2015.09.011

Hurley, D. H., and Benna, M. (2018). Simulations of lunar exospheric water events from meteoroid impacts. *Planetary Space Sci.*, doi: 10.1016/j.pss.2017.07.008

Ireland, T. R., Holden, P., Norman, M. D., and Clarke, J. (2006). Isotopic enhancements of [17]O and [18]O from solar wind particles in the lunar regolith. *Nature Lett.*, doi: 10.1038/nature04611

Johnstone, C. P., Lammer, H., Kislyakova, K. G., Scherf, M., and Güdel, M. (2021). The young Sun's XUV-activity as a constraint for lower $CO_2$-limits in the Earth's Archean atmosphere. *Earth Planet. Sci. Lett.*, 576, 117197, doi: 10.1016/j.epsl.2021.117197

Jones, B. M., Aleksandrov, A., Hibbitts, K., Dyar, M. D., and Orlando, T. M. (2018). Solar wind-induced water cycle on the Moon. *Geophysical Res. Lett.*, doi: 10.1029/2018GL080008

Kaiser, M. L., and Alexander, J. K. (1976). Source location measurements of terrestrial kilometric radiation obtained from lunar orbit. *Geophys. Res. Lett.*, doi: 10.1029/GL003i001p00037

Kallio, E., and Facskó, G. (2015). Properties of plasma near the moon in the magnetotail. *Planet. Space Sci.*, doi: 10.1016/j.pss.2014.11.007

Kallio, E., S. Dyadechkin, P. Wurz, and M. Khodachenko (2019). Space weathering on the Moon: Farside-nearside solar wind precipitation asymmetry. *Planet. Space Sci.*, 166, 9–22, doi: 10.1016/j.pss.2018.07.013

Kislyakova, K. G., Johnstone, C. P., Scherf, M., Holmström, M., Alexeev, I. I., Lammer, H., Khodachenko, M. L., and Güdel, M. (2020). Evolution of the Earth's polar outflow from mid-Archean to present. *J. Geophys. Res.*, doi: 10.1029/2020JA027837

Krimigis, S. M., Mitchell, D. G., Hamilton, D. C., Livi, S., Dandouras, J., Jaskulek, S., et al. (2004). Magnetospheric Imaging Instrument (MIMI) on the Cassini Mission to Saturn/Titan. *Space Sci. Rev.*, 114, doi: 10.1007/s11214-004-1410-8

Kronberg, E. A., Daly, P. W., Grigorenko, E. E., Smirnov, A. G., Klecker, B., and Malykhin, A. Y. (2021). Energetic charged particles in the terrestrial magnetosphere: Cluster/RAPID results. *J. Geophys. Res.*, doi: 10.1029/2021JA029273

Lammer, H., Kasting, J. F., Chassefière, E., Johnson, R. E., Kulikov, Y. N., and Tian, F. (2008). Atmospheric Escape and Evolution of Terrestrial Planets and Satellites. *Space Sci. Rev.*, doi: 10.1007/s11214-008-9413-5

Lammer, H., Zerkle, A. L., Gebauer, S., Tosi, N., Noack, L., Scherf, M., Pilat-Lohinger, E., Güdel, M., Grenfell, J. K., Godolt, M., and Nikolaou, A. (2018). Origin and evolution of the atmospheres of early Venus, Earth and Mars. *Astron. Astrophys. Rev.*, 26:2, doi: 10.1007/s00159-018-0108-y

Lammer, H., Scherf, M., Ito, Y., Mura, A., Vorburger, A., Guenther, E., Wurz, P., Erkaev, N. V., and Odert, P. (2022). The exosphere as a boundary: Origin and evolution of airless bodies in the inner solar system and beyond including planets with silicate atmospheres. *Space Sci. Rev.*, 218:15, doi: 10.1007/s11214-022-00876-5

Leblanc, F., and Chaufray, J. Y. (2011). Mercury and Moon He exospheres: Analysis and modelling. *Icarus*, doi: 10.1016/j.icarus.2011.09.028

Leblanc, F., Schmidt, C., Mangano, V. et al. (2022). Comparative Na and K Mercury and Moon Exospheres. *Space Sci. Rev.*, 218, doi: 10.1007/s11214-022-00871-w

Lebreton, J.-P., Stverak, S., Travnicek, P., Maksimovic, M., et al. (2006). The ISL Langmuir probe experiment processing onboard DEMETER: Scientific objectives, description and first results. *Planet. Space Sci.*, 54, doi: 10.1016/j.pss.2005.10.017

Lefeuvre, F., Blanc, E., Pinçon, J. L., et al. (2008). TARANIS - A Satellite Project Dedicated to the Physics of TLEs and TGFs. *Space Sci. Rev.*, 137, doi: 10.1007/s11214-008-9414-4

Liu, Y., Guan, Y., Zhang, Y., et al. (2012). Direct measurement of hydroxyl in the lunar regolith and the origin of lunar surface water. *Nature Geosci.*, doi: 10.1038/ngeo1601

Looper, M. D., et al., (2013). The radiation environment near the lunar surface: CRaTER observations and Geant4 simulations. *Space Weath.*, 11, doi: 10.1002/swe.20034

Marty, B., Hashizume, K., Chaussidon, M., and Wieler, R. (2003). Nitrogen isotopes on the Moon: archives of the solar and planetary contributions to the inner solar system. *Space Sci. Rev.*, 106, 175–196, doi: 10.1023/A:1024689721371

McComas, D. J., Allegrini, F., Bochsler, P., et al. (2009a). Lunar backscatter and neutralization of the solar wind: First observations of neutral atoms from the Moon. *Geophys. Res. Lett.*, 36, doi: 10.1029/2009GL038794

McComas, D. J., Allegrini, F., Baldonado, J., et al. (2009b). The Two Wide-angle Imaging Neutral-atom Spectrometers (TWINS) NASA Mission-of-Opportunity. *Space Sci. Rev.*, 142, doi: 10.1007/s11214-008-9467-4

McCord, T. B., Taylor, L. A., Combes, J-P., Kramer, G., Pieters, C. M., Sunshine, J. M., and Clark, R. N. (2011). Sources and physical processes responsible for $OH/H_2O$ in the lunar soil as revealed by the Moon Mineralogy Mapper ($M^3$). *J. Geophys. Res.*, 116, doi: 10.1029/2010JE003711

McLain, J. L., Loeffler, M. J., Farrell, W. M., Honniball, C. I., Keller, J. W., and Hudson, R. (2021). Hydroxylation of Apollo 17 soil sample 78421 by solar wind protons. *J. Geophys. Res.*, 126, doi: 10.1029/2021JE006845

Mewaldt, R. A., Leske, R. A., Stone, E. C., Barghouty, A. F., Labrador, A. W., Cohen, C. M. S., Cummings, A. C., Davis, A. J., von Rosenvinge, T. T., and Wiedenbeck, M. E. (2009). STEREO Observations of Energetic Neutral Hydrogen Atoms During the 2006 December 5 Solar Flare. *Astroph. J.*, doi: 10.1088/0004-637X/693/1/L11

Mitchell, D., Jaskulek, S., Schlemm, C. et al. (2000). High energy neutral atom (HENA) imager for the IMAGE mission. *Space Sci. Rev.*, 91, doi: 10.1023/A:1005207308094

Mitchell, D. L., Mazelle, C., Sauvaud, J. A. et al. (2016). The MAVEN Solar Wind Electron Analyzer. *Space Sci. Rev.*, 200, doi: 10.1007/s11214-015-0232-1

Mrigakshi, A. I., Matthiä, D., Berger, T., Reitz, G., and Wimmer-Schweingruber, R. F. (2012). Assessment of galactic cosmic ray models. *J. Geophys. Res.*, 117, doi: 10.1029/2012JA017611

Nagai, T., Tsunakawa, H., Shibuya, H., Takahashi, F., Shimizu, H., Matsushima, M., Nishino, M. N., Yokota, Y., Asamura, K., Tanaka, T., Saito, Y., and Amm, O. (2009). Plasmoid formation for multiple onset substorms: observations of the Japanese Lunar Mission "Kaguya". *Ann. Geophys.*, 27, doi: 10.5194/angeo-27-59-2009

Nakamura, R. (2006). Substorms and Their Solar Wind Causes. *Space Sci. Rev.*, doi: 10.1007/s11214-006-9131-9

Nénon, Q., and Poppe, A. R. (2020). On the Long-term Weathering of Airless Body Surfaces by the Heavy Minor Ions of the Solar Wind: Inputs from Ion Observations and SRIM Simulations. *Planet. Sci. J.*, doi: 10.3847/PSJ/abbe0c

Nénon, Q., and Poppe, A. R. (2021). Bombardment of Lunar Polar Crater Interiors by Out-of-ecliptic Ions: ARTEMIS Observations. *Planet. Sci. J.*, doi: 10.3847/PSJ/abfda2

Nishino, M. N., et al. (2010). Effect of the solar wind proton entry into the deepest lunar wake. *Geophys. Res. Lett.*, doi: 10.1029/2010GL043948

Orsini, S., Livi, S.A., Lichtenegger, H. et al. (2021). SERENA: Particle Instrument Suite for Determining the Sun-Mercury Interaction from BepiColombo. *Space Sci. Rev.*, 217, doi: 10.1007/s11214-020-00787-3

Ozima, M., Seki, K., N. Terada, K., Miura, Y. N., Podosek, F. A., and Shinagawa, H. (2005). Terrestrial nitrogen and noble gases in lunar soils. *Nature*, doi: 10.1038/nature03929

Parks, G., Chen, L., Fillingim, M., et al. (2001). Kinetic Characterization of Plasma Sheet Dynamics. *Space Sci. Rev.*, 95, doi: 10.1023/A:1005206701965

Pieters, C. M., and Noble, S. K. (2016). Space weathering on airless bodies. *J. Geophys. Res.*, doi: 10.1002/2016JE005128

Plainaki, C., Lilensten, J., Radioti, A., Andriopoulou, M., Milillo, A., Nordheim, T. A., Dandouras, I., Coustenis, A., Grassi, D., Mangano, V., Massetti, S., Orsini, S., and Lucchetti, A. (2016). Planetary space weather: scientific aspects and future perspectives. *J. Space Weather Space Clim.*, doi: 10.1051/swsc/2016024

Popel, S. I., Zelenyi, L. M., Golub, A. P., and Dubinskii, A. Yu. (2018). Lunar dust and dusty plasmas: Recent developments, advances, and unsolved problems. *Planet. Space Sci.*, doi: 10.1016/j.pss.2018.02.010

Popel, S. I., Golub, A. P., Kassem, A. I., and Zelenyi, L. M. (2022). Dust dynamics in the lunar dusty plasmas: Effects of magnetic fields and dust charge variations. *Physics of Plasmas*, doi: 10.1063/5.0077732

Poppe, A. R., Samad, R., Halekas, J. S., Sarantos, M., Delory, G. T., Farrell, W. M., Angelopoulos, V., and McFadden, J. P. (2012a). ARTEMIS observations of lunar pick-up ions in the terrestrial magnetotail lobes. *Geophys. Res. Lett.*, doi: 10.1029/2012GL052909

Poppe, A. R., Piquette, M., Likhanskii, A., and Horányi, M. (2012b). The effect of surface topography on the lunar photoelectron sheath and electrostatic dust transport. *Icarus*, doi: 10.1016/j.icarus.2012.07.018

Poppe, A. R., Fatemi, S., Garrick-Bethell, I., Hemingway, D., and Holmström, M. (2015). Solar wind interaction with the Reiner Gamma crustal magnetic anomaly: Connecting source magnetization to surface weathering. *Icarus*, doi: 10.1016/j.icarus.2015.11.005

Poppe, A. R., Fillingim, M. O., Halekas, J. S., Raeder, J., and Angelopoulos, V. (2016). ARTEMIS observations of terrestrial ionospheric molecular ion outflow at the Moon. *Geophys. Res. Lett.*, doi: 10.1002/2016GL069715

Poppe, A. R., Halekas, J. S., and Harada, Y. (2022). A comprehensive model for pickup ion formation at the Moon. *J. Geophys. Res.*, 127, doi: 10.1029/2022JE007422

Potter, A. E., Killen, R. M., and Morgan, T. H. (2000). Variation of lunar sodium during passage of the Moon through the Earth's magnetotail. *J. Geophys. Res.*, doi: 10.1029/1999JE001213

Quinn, P. R., Schwadron, N. A., Townsend, L. W., Wimmer-Schweingruber, R. F., Case, A. W., Spence, H. E., Wilson J. K., and Joyce, C. J. (2017). Modeling the effectiveness of shielding in the earth-moon-mars radiation environment using PREDICCS: five solar events in 2012. *J. Space Weather Space Clim.*, doi: 10.1051/swsc/2017014

Rairden, R. L., Frank, L. A., and Craven, J. D. (1986). Geocoronal imaging with Dynamics Explorer. *J. Geophys. Res.*, doi: 10.1029/JA091iA12p13613

Ratti, F., Stankov A., Wirth, K., Agnolon, V., Rando, N., Corral, C., Kueppers, M., Asquier, J., Ertel H., and Wielders, A. (2022). Instruments on board the Comet Interceptor ESA mission. In*: Proc. 4S Symposium 2022*

Rème, H., Aoustin, C., Bosqued, J.M., Dandouras, I., et al. (2001). First multispacecraft ion measurements in and near the Earth's magnetosphere with the identical Cluster ion spectrometry (CIS) experiment. *Ann. Geophys.*, doi: 10.5194/angeo-19-1303-2001

Robertson, I. P., et al. (2009). Solar wind charge exchange observed through the lunar exosphere. *Geophys. Res. Lett.*, 36, doi: 10.1029/2009GL040834

Šafránková, J., Němeček, Z., Přech, L. et al. (2013). Fast Solar Wind Monitor (BMSW): Description and First Results. *Space Sci. Rev.*, 175, doi: 10.1007/s11214-013-9979-4

Saito, Y., et al. (2010). In-flight Performance and Initial Results of Plasma Energy Angle and Composition Experiment (PACE) on SELENE (Kaguya). *Space Sci. Rev.*, doi: 10.1007/s11214-010-9647-x

Sandel, B. R., Goldstein, J., Gallagher, D. L., and Spasojević, M. (2003). Extreme ultraviolet imager observations of the structure and dynamics of the plasmasphere. *Space Sci. Rev.*, 109, doi: 10.1023/B:SPAC.0000007511.47727.5b

Sarrailh, P., Matéo-Vélez, J.-C., Hess, S. L. G., Roussel, J.-F., Thiébault, B., Forest, J., Jeanty-Ruard, B., Hilgers, A., Rodgers, D., Cipriani, F., and Payan D. (2015). SPIS 5: New Modeling Capabilities and Methods for Scientific Missions. *IEEE Transactions on Plasma Science*, 43 (9), 2789–2798, doi: 10.1109/TPS.2015.2445384

Sauvaud, J. A., Moreau, T., Maggiolo, R., Treilhou, J.-P., et al. (2006). High-energy electron detection onboard DEMETER: The IDP spectrometer, description and first results on the inner belt. *Planetary and Space Science*, doi: 10.1016/j.pss.2005.10.019

Schörghofer, N., Benna, M., Berezhnoy, A. A., Greenhagen, B., Jones, B. M., Li, S., Orlando, T. M., Prem, P., Ticker, O. J., and Wöhler, C. (2021). Water group exospheres and surface interactions on the Moon, Mercury, and Ceres. *Space Sci. Rev.*, 217, doi: 10.1007/s11214-021-00846-3

Schwadron, N. A., Wilson, J. K., Looper, M. D., et al. (2016). Signatures of volatiles in the lunar proton albedo. *Icarus*, doi: 10.1016/j.icarus.2015.12.003

Schwadron, N. A., Wilson, J. K., Jordan, A. P., et al. (2018). Using proton radiation from the moon to search for diurnal variation of regolith hydrogenation. *Planetary and Space Science*, 162, doi: 10.1016/j.pss.2017.09.012

Seki, K., Terasawa, T., Hirahara, M., and Mukai, T. (1998). Quantifications of tailward cold $O^+$ beams in the lobe/mantle with Geotail data: Constraints on polar $O^+$ outflows. *J. Geophys. Res.*, 103(A12), 29371–29382, doi: 10.1029/98JA02463

Sibeck, D. G., Allen, R., Aryan, H. et al. (2018). Imaging Plasma Density Structures in the Soft X-Rays Generated by Solar Wind Charge Exchange with Neutrals. *Space Sci Rev.*, 214, doi: 10.1007/s11214-018-0504-7

Sitnov, M., Birn, J., Ferdousi, B., et al. (2019). Explosive Magnetotail Activity. *Space Sci. Rev.*, 215, doi: 10.1007/s11214-019-0599-5

Sohn, J., Oh, S., and Yi, Y. (2014). Lunar cosmic ray radiation environments during Luna and Lunar Reconnaissance Orbiter missions. *Adv. Space Res.*, doi: 10.1016/j.asr.2014.05.011

Stern, S. A. (1999). The lunar atmosphere: History, status, current problems, and context. *Rev. Geophys.*, 37 (4), 453–491, doi: 10.1029/1999RG900005

Stern, S. A., Cook, J. C., Chaufray, J-Y., Feldman, P. D., Gladstone, G. R., and Retherford, K. D. (2013). Lunar atmospheric $H_2$ detections by the LAMP UV spectrograph on the Lunar Reconnaissance Orbiter. *Icarus*, 226, 1210–1213, doi: 10.1016/j.icarus.2013.07.011

Stubbs, T. J., Vondrak, R. R., and Farrell, W. M. (2007). A dynamic fountain model for dust in the lunar exosphere. *Dust in Planetary Systems*, ESA SP-643

Stubbs, T. J., Farrell, W. M., Halekas, J. S., Burchill, J. K., et al. (2013). Dependence of lunar surface charging on solar wind plasma conditions and solar irradiation. *Planetary Space Sci.,* doi: 10.1016/j.pss.2013.07.008

Taylor, M. G. G. T., et al. (2006). Cluster encounter with an energetic electron beam during a substorm. *J. Geophys. Res.*, doi: 10.1029/2006JA011666

Terada, K., Yokota, S., Saito, Y., Kitamura, N., Asamura, K., and Nishino, M. N. (2017). Biogenic oxygen from Earth transported to the Moon by a wind of magnetospheric ions. *Nature Astronomy*, doi: 10.1038/s41550-016-0026

Tucker, O. J., Farrell, W. M., Killen, R. M., and Hurley, D. M. (2019). Solar Wind Implantation into the Lunar Regolith: Monte Carlo Simulations of H Retention in a Surface with Defects and the H2 Exosphere. *J. Geophys. Res.*, doi: 10.1029/2018JE005805

Vallat, C., Dandouras, I., C:son Brandt, P., Mitchell, D. G., Roelof, E. C., deMajistre, R., Rème, H., Sauvaud, J.-A., Kistler, L., Mouikis, C., Dunlop, M., and Balogh, A. (2004). First comparisons of local ion measurements in the inner magnetosphere with ENA magnetospheric image inversions: Cluster-CIS and IMAGE-HENA observations. *J. Geophys. Res.*, doi: 10.1029/2003JA010224

von Steiger, R. (2008). The SolarWind Throughout the Solar Cycle. In *The Heliosphere through the Solar Activity Cycle*, Springer Praxis Books, Chapter 3, doi: 10.1007/978-3-540-74302-6_3

Vorburger, A., Wurz, P., Barabash, S., Wieser, M., Futaana, Y., Lue, C., Holmström, M., Bhardwaj, A., Dhanya, M. B., and Asamura, K. (2013). Energetic neutral atom imaging of the lunar surface. *J. Geophys. Res.*, doi: 10.1002/jgra.50337

Vorburger, A., Wurz, P., Barabash, S., Wieser, M., Futaana, Y., Bhardwaj, A., and Asamura, K. (2015). Imaging the South Pole–Aitken basin in backscattered neutral hydrogen atoms. *Planet. Space Sci.*, doi: 10.1016/j.pss.2015.02.007

Vorburger, A., Wurz, P., Barabash, S., Futaana, Y., Wieser, M., Bhardwaj, A., Dhanya, M. B., and Asamura, K. (2016). Transport of solar wind plasma onto the lunar nightside surface. *Geophys. Res. Lett.*, doi: 10.1002/2016GL071094

Wang, H. Z., et al. (2021). Earth Wind as a Possible Exogenous Source of Lunar Surface Hydration. *Astrophys. J. Lett.*, doi: 10.3847/2041-8213/abd559

Wang, X.-D., Zong, Q.-G., Wang, J.-S., Cui, J., Rème, H., Dandouras, I., et al. (2011), Detection of m/q = 2 pickup ions in the plasma environment of the Moon: The trace of exospheric H2+. *Geophys. Res. Lett.*, doi: 10.1029/2011GL047488

Wang, X.-Q., Cui, J., Wang, X.-D., Liu, J.-J., Zhang, H.-B., Zuo, W., Su, Y., Wen, W.-B., Rème, H., Dandouras, I., Aoustin, C., Wang, M., Tan, X., Shen, J., Wang, F., Fu, Q., Li, C.-L., and Ouyang, Z.-Y. (2012). The Solar Wind interactions with Lunar Magnetic Anomalies: A case study of the Chang'E-2 plasma data near the Serenitatis antipode. *Adv. Space Res.*, doi: 10.1016/j.asr.2011.12.003

Wei, Y., Zhong, J., Hui, H., Shi, Q., Cui, J., He, H., Zhang, H., Yao, Z., Yue, X., Rong, Z., He, F., Chai, L., and Wan, W. (2020). Implantation of Earth's atmospheric ions into the nearside and farside lunar soil: implications to geodynamo evolution. *Geophys. Res. Lett.*, 47, e2019GL086208, doi: 10.1029/2019GL086208

Whitley, R., and Martinez, R. (2016). Options for staging orbits in cislunar space. *2016 IEEE Aerospace Conference*, 1–9, doi: 10.1109/AERO.2016.7500635

Wieser, M., Barabash, S., Futaana, Y., Holmström, M., Bhardwaj, A., Sridharan, R., Dhanya, M. B., Wurz, P., Schaufelberger, A., and Asamura , K. (2009). Extremely high reflection of solar wind protons as neutral hydrogen atoms from regolith in space. *Planet. Space Sci.*, doi: 10.1016/j.pss.2009.09.012

Wieser, M., Barabash, S., Futaana, Y., Holmström, M., Bhardwaj, A., Sridharan, R., Dhanya, M. B., Schaufelberger, A., Wurz, P., and Asamura, K. (2010). First observation of a mini-magnetosphere above a lunar magnetic anomaly using energetic neutral atoms. *Geophys. Res. Lett.*, doi: 10.1029/2009GL041721

Wu, X., Ambrosi, G., Azzarello, P., Bergmann, B., et al. (2019). Penetrating particle ANalyzer (PAN). *Adv. Space Res.*, doi: 10.1016/j.asr.2019.01.012

Wurz, P., Rohner, U., Whitby, J. A., Kolb, C., Lammer, H., Dobnikar, P., and Martín-Fernández, J.A. (2007). The lunar exosphere: The sputtering contribution. *Icarus*, doi: 10.1016/j.icarus.2007.04.034

Wurz, P., Fatemi, S., Galli, A. et al. (2022). Particles and Photons as Drivers for Particle Release from the Surfaces of the Moon and Mercury. *Space Sci. Rev.*, doi: 10.1007/s11214-022-00875-6

Yokota, S., et al. (2009). First direct detection of ions originating from the Moon by MAP-PACE IMA onboard SELENE (KAGUYA). *Geophys. Res. Lett.*, 36, doi: 10.1029/2009GL038185

Zahnle, K. J., Gacesa, M., and Catling, D. C. (2019). Strange messenger: a new history of hydrogen on Earth, as told by Xenon. *Geochim. Cosmochim. Acta*, 244, 56–85, doi: 10.1016/j.gca.2018.09.017

Zaman, F. A., Townsend, L. W., de Wet, W. C., Looper, M. D., Brittingham, J. M., Burahmah, N. T., et al. (2022). Modeling the lunar radiation environment: a comparison among FLUKA, Geant4, HETC-HEDS, MCNP6, and PHITS. *Space Weather*, doi: 10.1029/2021SW002895

Zhang, S., Wimmer-Schweingruber, R. F., Yu, J., Wang, C., et al. (2020). First measurements of the radiation dose on the lunar surface. *Sci. Adv.*, doi: 10.1126/sciadv.aaz1334

Zoennchen, J. H., Nass, U., Fahr, H. J., and Goldstein, J. (2017). The response of the H geocorona between 3 and 8 Re to geomagnetic disturbances studied using TWINS stereo Lyman-α data. *Ann. Geophys.*, doi: 10.5194/angeo-35-171-2017

Zoennchen, J. H., Connor, H. K., Jung, J., Nass, U., and Fahr, H. J. (2022). Terrestrial exospheric dayside H-density profile at $3–15R_E$ from UVIS/HDAC and TWINS Lyman-α data combined. *Ann. Geophys.*, doi: 10.5194/angeo-40-271-2022

# 1    Data Availability Statement

N/A.

# 2    Figures

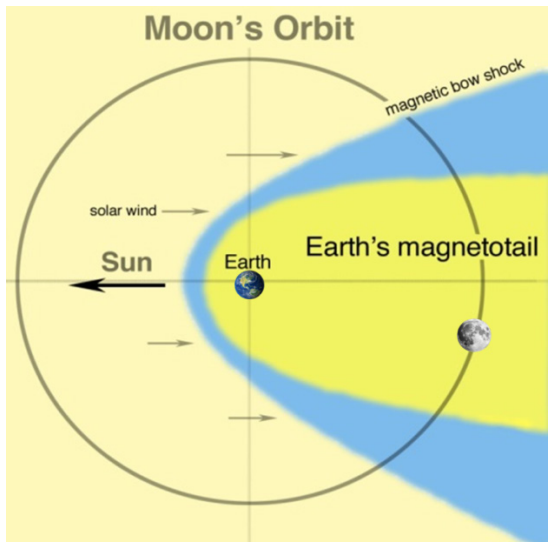

**FIGURE 1 |** Moon's orbit with respect to the Earth's magnetosphere. Earth's and Moon's sizes are not on scale. (Adapted from: Tim Stubbs / University of Maryland / GSFC).

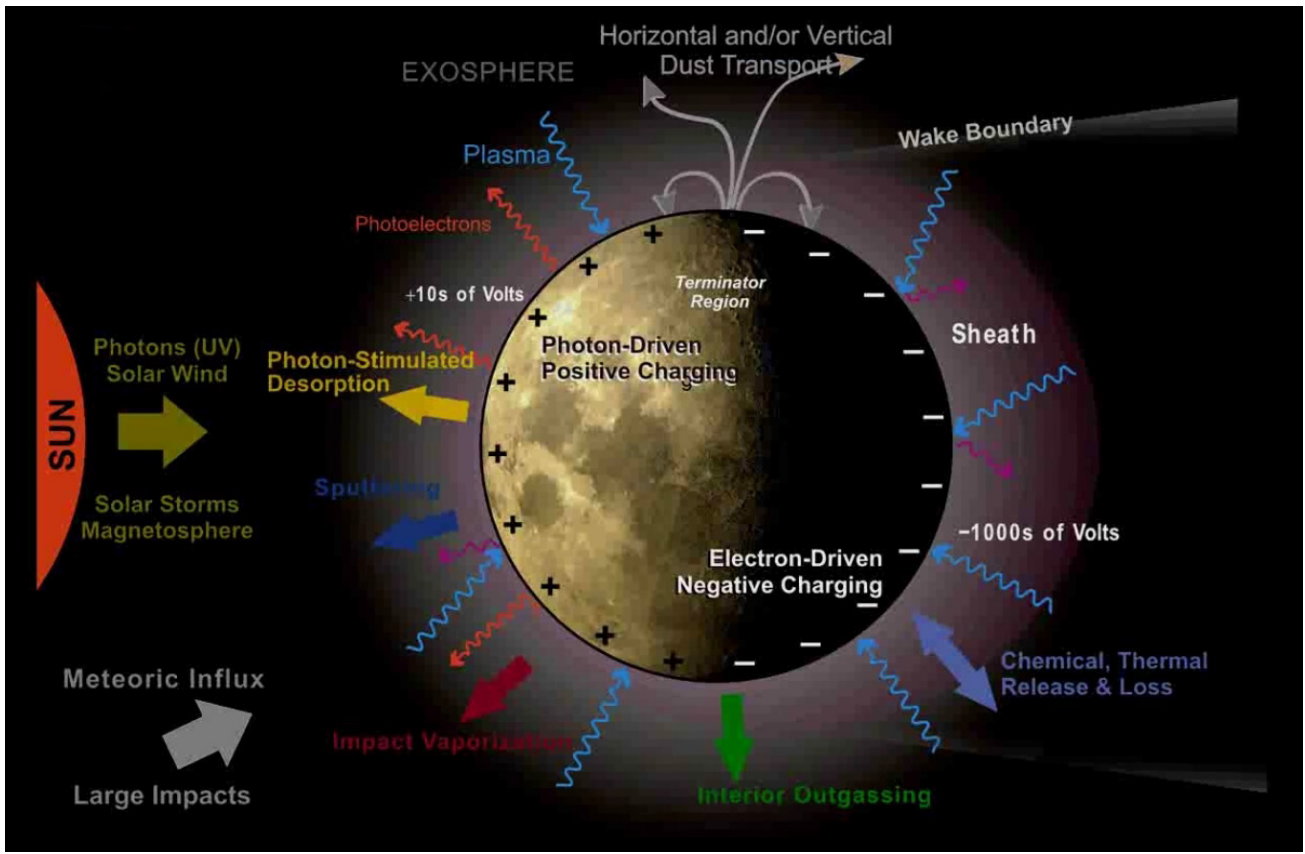

**FIGURE 2 |** Moon's environment with the complex interaction between solar radiation, space plasma, meteoritic flux, dust, exosphere and the surface (Credit: Jasper Halekas).

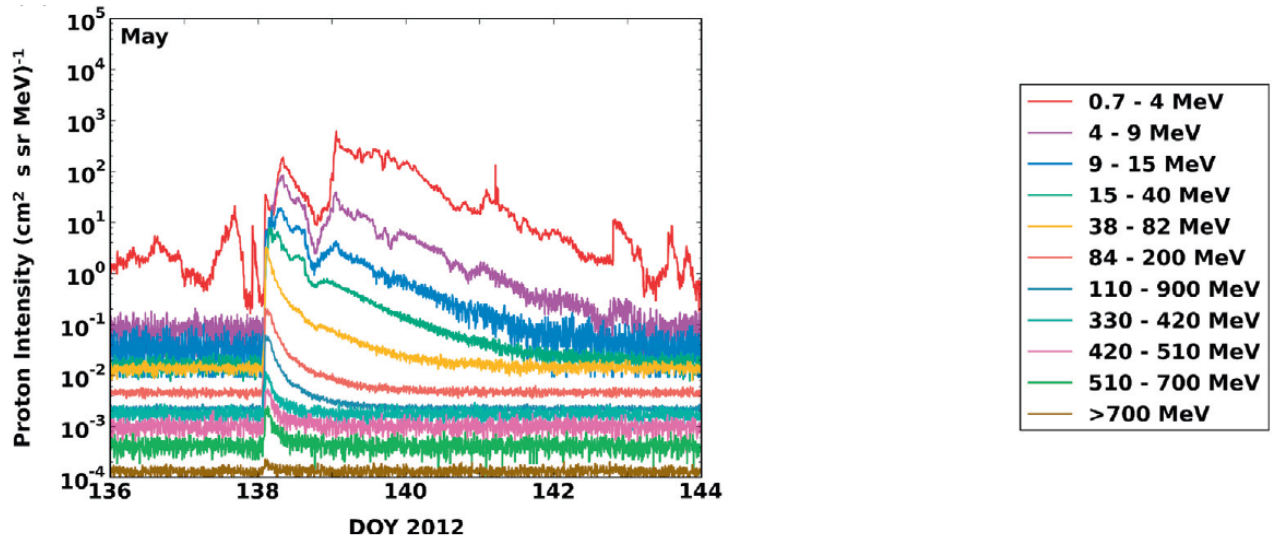

**FIGURE 3 |** Typical SEP (Solar Energetic Particles) proton intensities: five-minute averages of proton intensities measured by GOES-13/EPS/HEPAD during the May 2012 solar events. (From: Quinn et al., 2017).

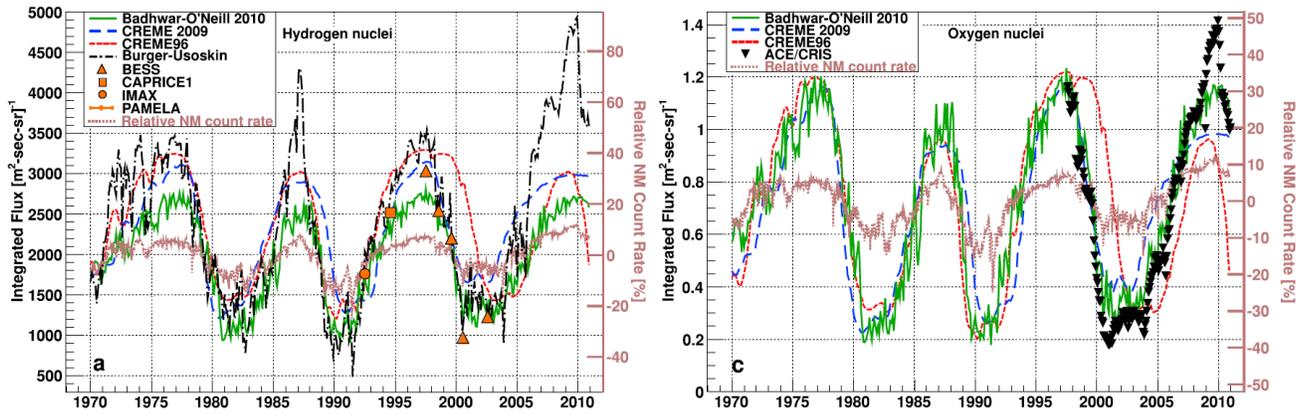

**FIGURE 4 |** Typical GCR (Galactic Cosmic Rays) Hydrogen nuclei (left) and Oxygen nuclei (right) fluxes. (From: Mrigakshi et al., 2012).

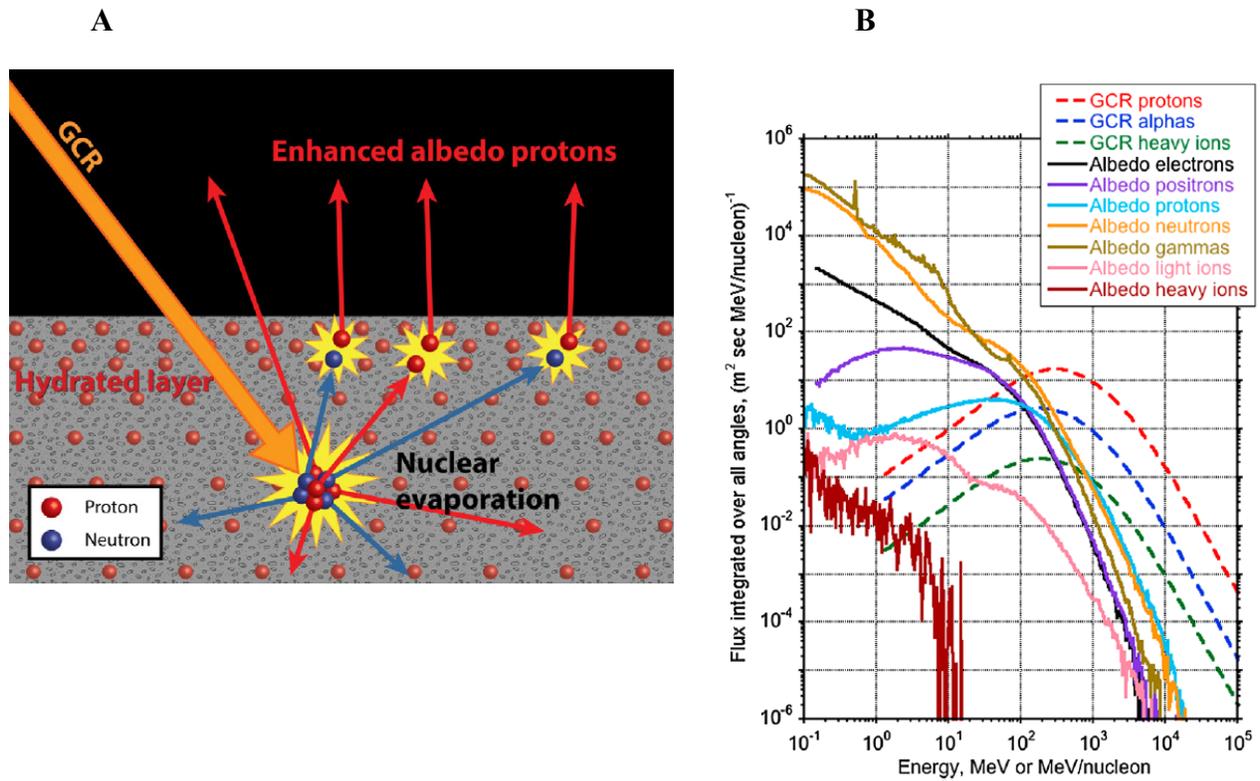

**FIGURE 5 | (A)**: Illustration of the effects of a hydrated layer of lunar regolith in the production of GCR albedo (secondary) protons. The nuclear evaporation process from deep in the regolith produces abundant secondary particles in all directions. (From: Schwadron et al., 2016). **(B)**: Energy spectra of pristine GCR species (dashed lines) and of lunar albedo species (continuous lines), calculated with the Geant4 simulation toolkit. (From: Looper et al., 2013).

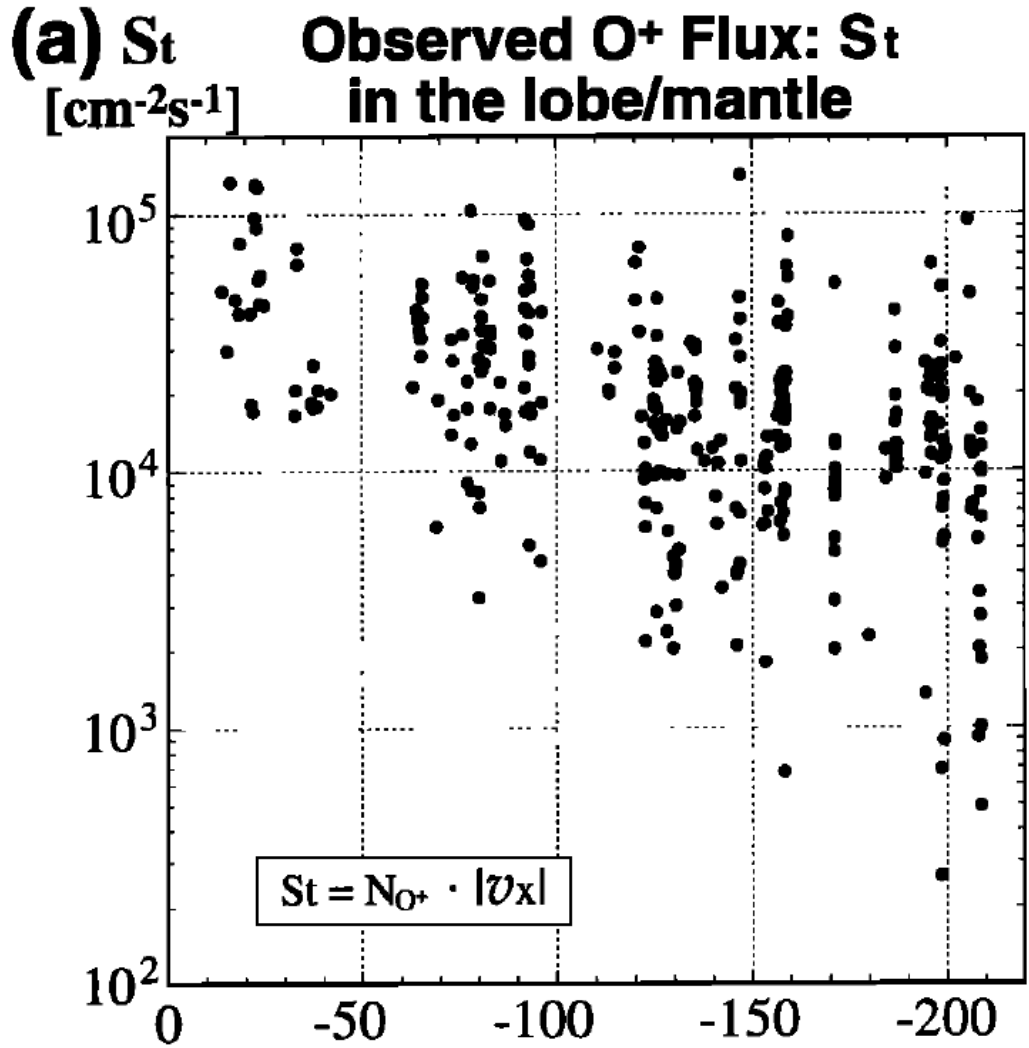

**FIGURE 6 |** Cold O$^+$ beam fluxes, observed by the Geotail spacecraft in the magnetotail lobe and plasma sheet boundary layer, versus the tailward distance from the Earth (X$_{GSM}$ in R$_E$). The Moon is at X$_{GSM}$ ≈ -60 R$_E$ (Earth radii). (From: Seki et al., 1998).

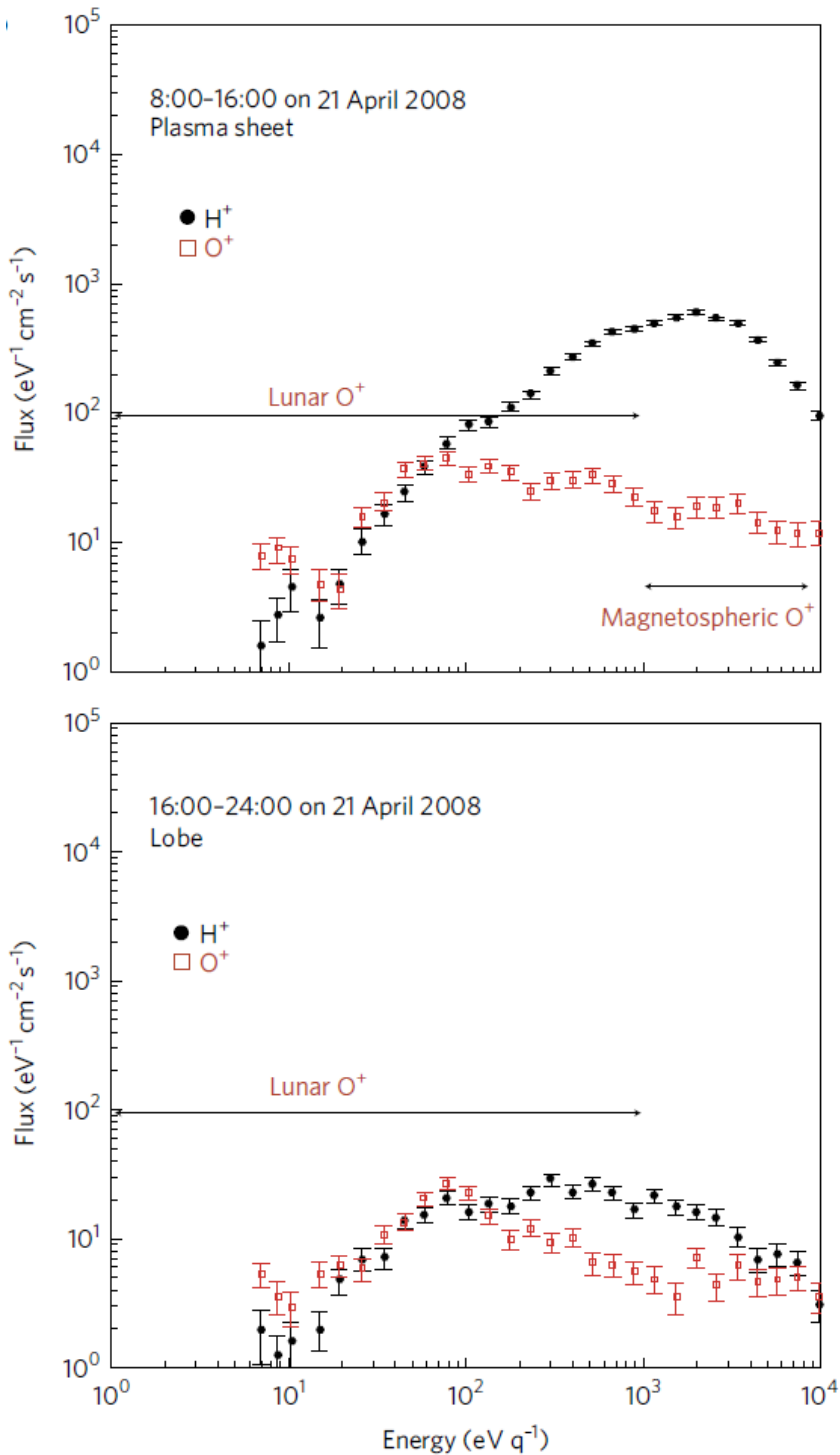

**FIGURE 7 |** Energy distributions of H+ and O+ ions measured by the IMA sensor onboard the Kaguya lunar orbiter in the terrestrial magnetotail. During the plasma sheet encounter (top panel) there is an enhancement of high-energy (1 – 10 keV) O+ ions, in comparison to those measured in the magnetotail lobe (bottom panel). The calculated density and net flux of these magnetospheric O+ ions, during the plasma sheet encounter, were $1.2 \times 10^{-3}$ cm$^{-3}$ and $2.6 \times 10^{4}$ cm$^{-2}$ s$^{-1}$ respectively. (From: Terada et al., 2017).

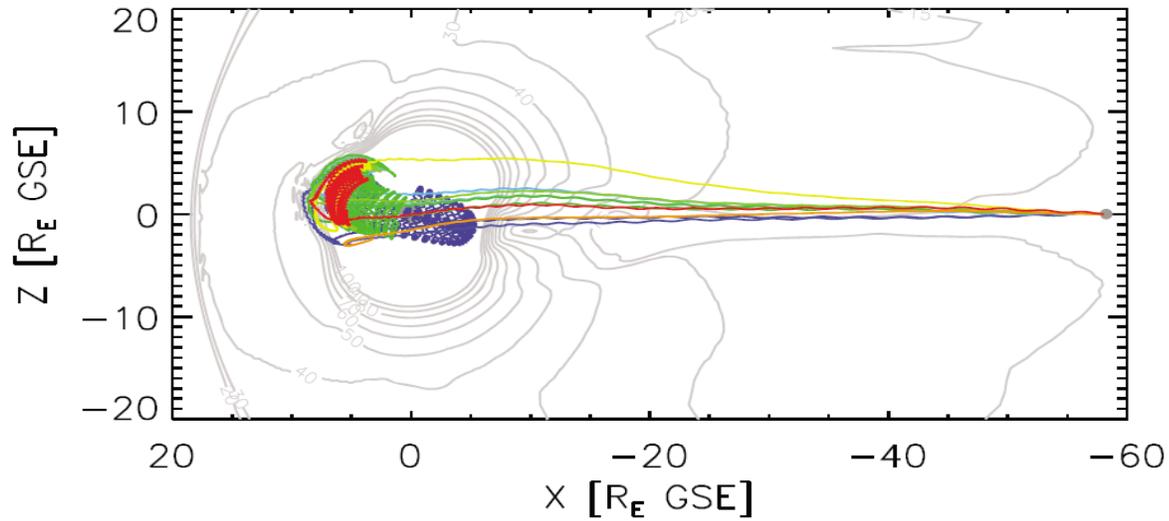

**FIGURE 8 |** MHD Open Global Geospace Circulation Model simulations (backward particle tracing) suggest how heavy ions, observed in the Moon environment during high geomagnetic activity events (at $X_{GSE} \approx$ -60 $R_E$), can be originating from the inner magnetosphere. Earth-to-Moon transport times are ~2 – 3 hours. (From: Poppe et al., 2016).

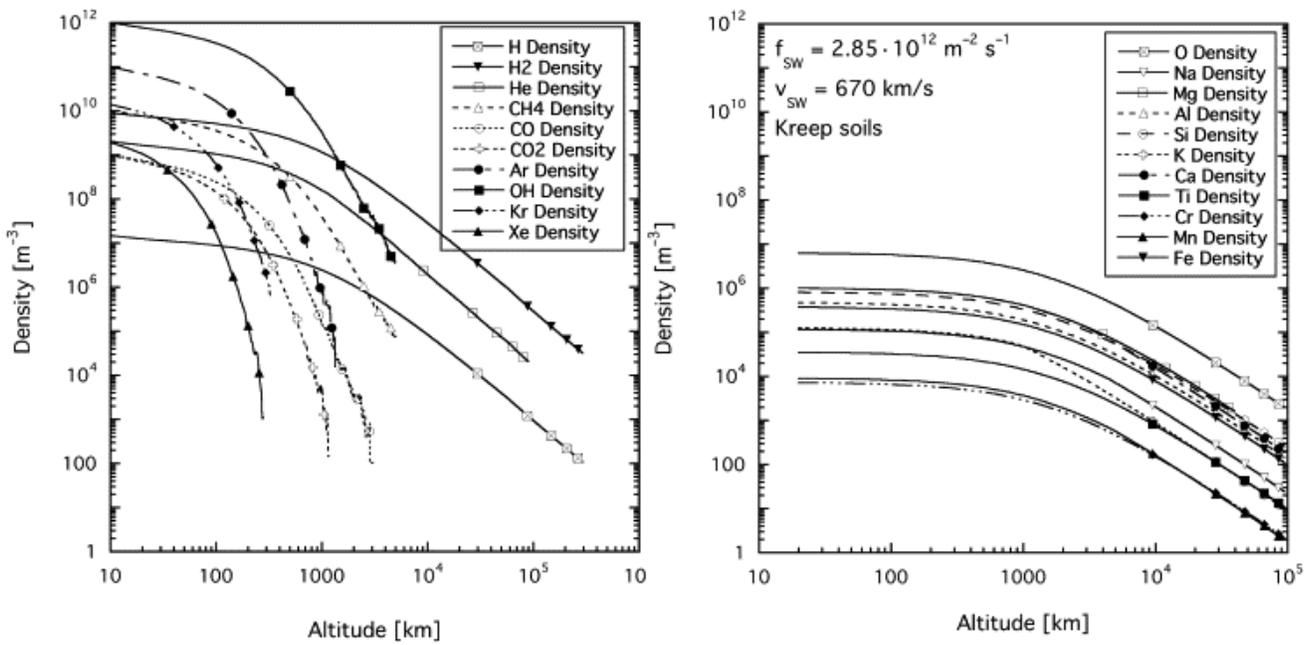

**FIGURE 9 |** Left: Lunar exosphere density profiles for the atoms and molecules thermally released from surface; based on the exospheric surface densities from Stern (1999). Right: Lunar exosphere density profiles for the atoms released through sputtering. Both calculations are done for the subsolar point. (From: Wurz et al., 2007).

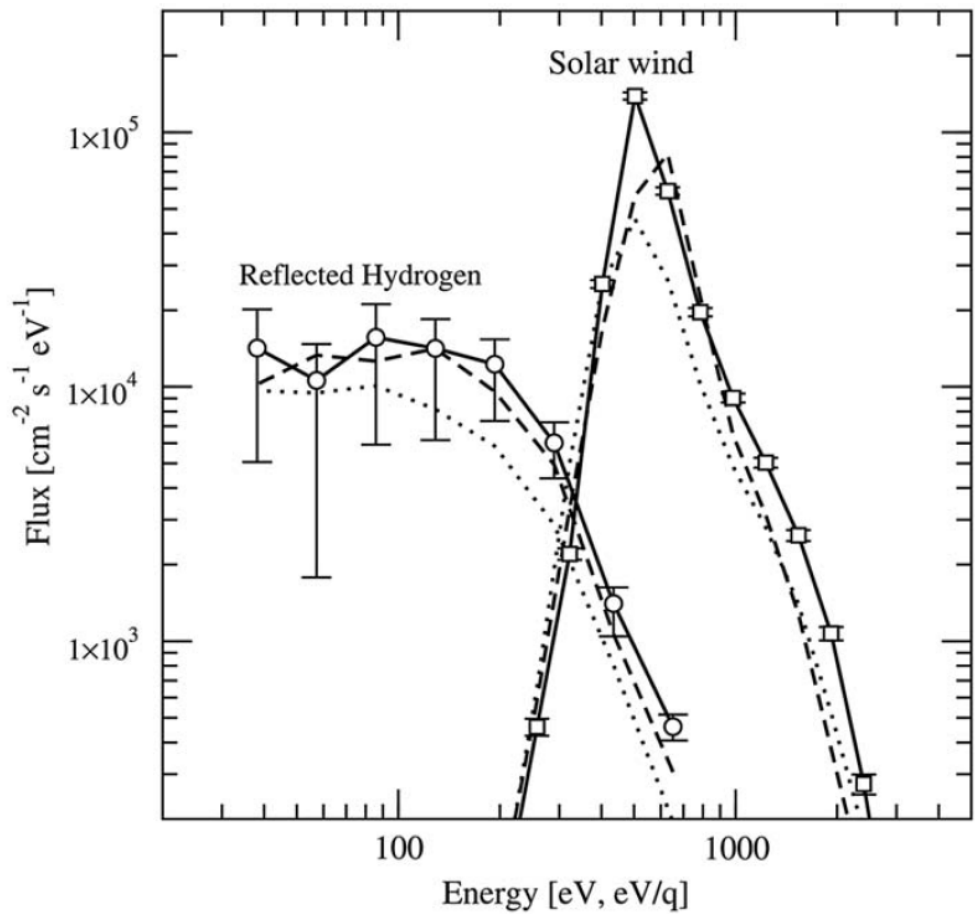

**FIGURE 10 |** Typical energy spectra of the solar wind ions (right side, open squares) and of the corresponding reflected from the lunar regolith energetic hydrogen atoms (left side, open circles), measured by the SARA instrument onboard the Chandrayaan-1 spacecraft. Note the good correlation between the reflected energetic neutral flux and the solar wind flux variations. (From: Wieser et al., 2009).

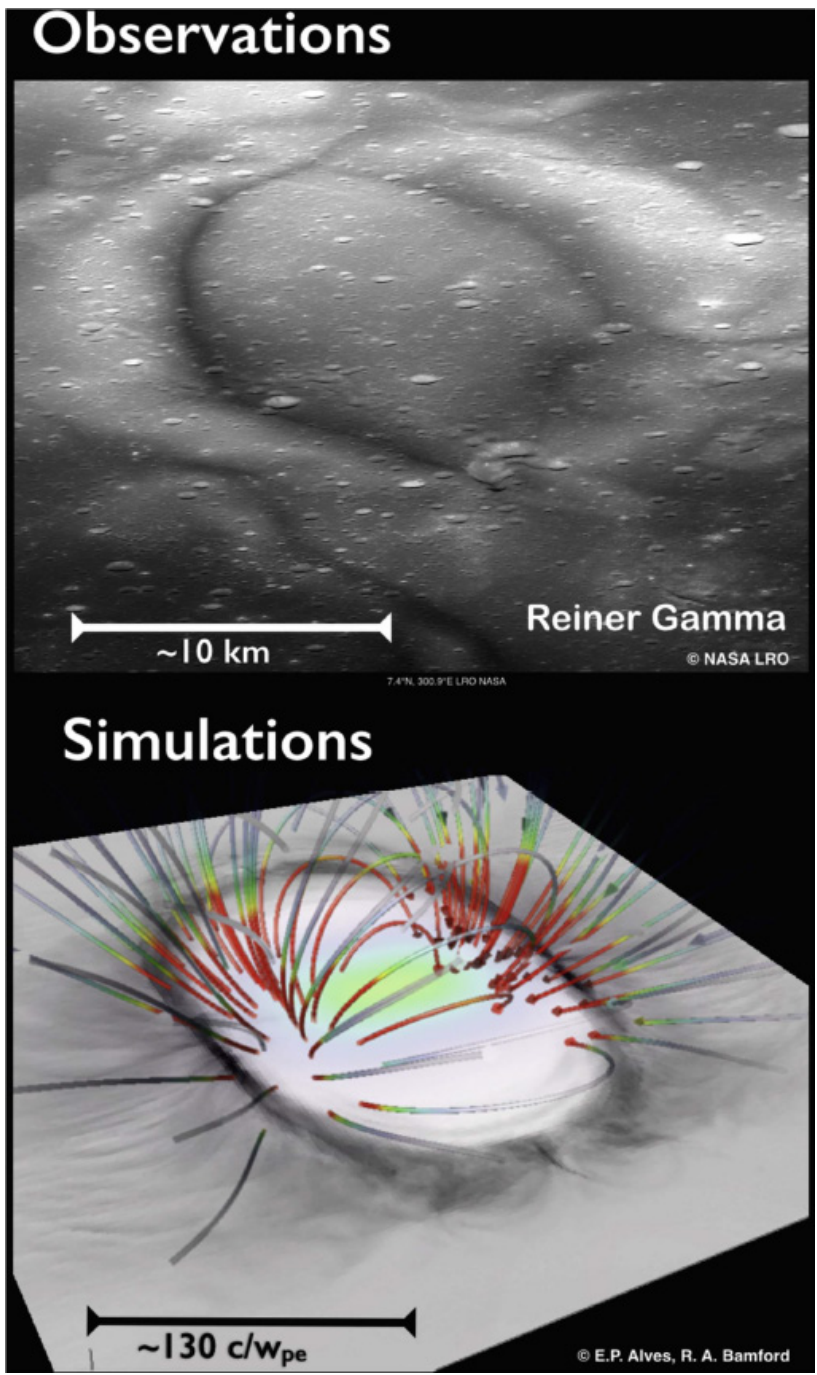

**FIGURE 11 |** Top: Image of the central region of the Reiner Gamma Formation lunar swirl, taken by Lunar Reconnaissance Orbiter. Bottom: A slice of the relative solar wind proton density above this lunar swirl obtained from a 3D simulation, with the initial magnetic field lines corresponding to a single subsurface dipole. (From: Bamford et al., 2016).

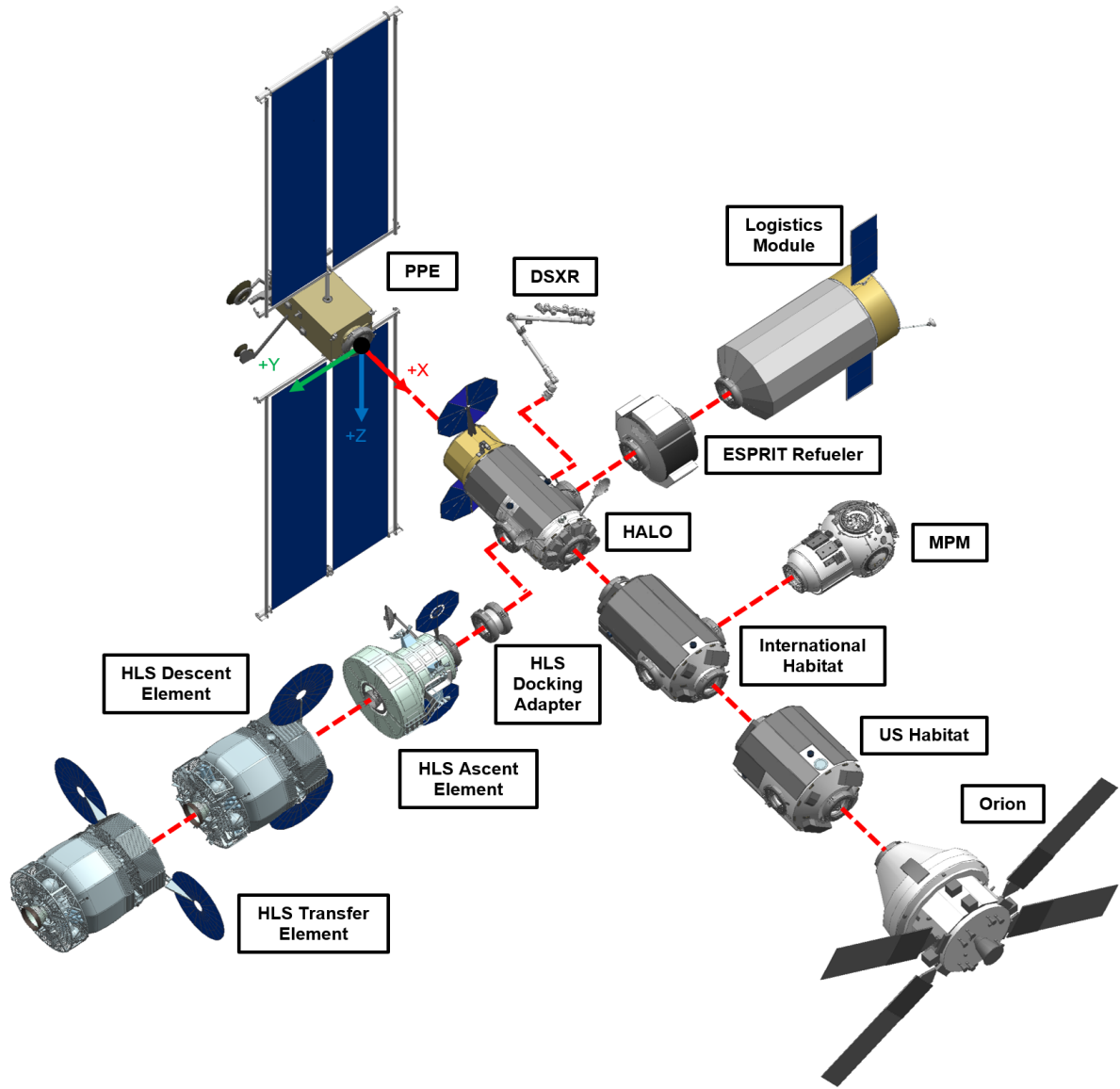

**FIGURE 12 |** Gateway configuration. PPE: Power and Propulsion Element; HLS: Human Landing System; HALO: Habitation and Logistics Outpost (Credit: NASA).

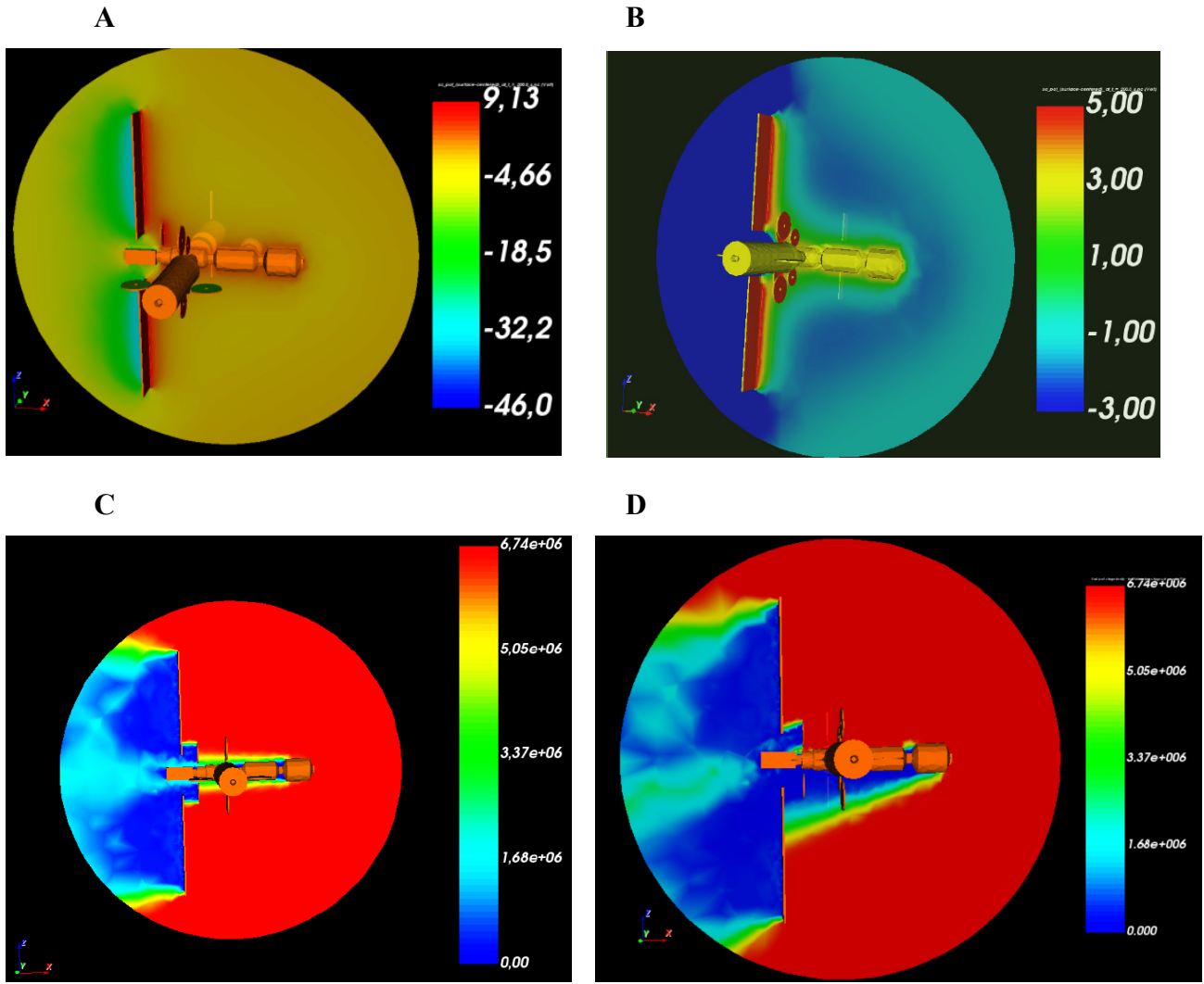

**FIGURE 13 |** Top row: SPIS (Spacecraft Plasma Interaction System) simulation of the volume electrostatic potential distribution (in V) in the solar wind, with the Gateway aligned to the solar direction. **(A)**: Full potential scale. **(B)**: Potential scale saturated at +5 / -3 V, to highlight the potential values away from the solar panels. Bottom row: Proton density (in m⁻³) in the solar wind, with the Gateway aligned to the solar direction **(C)** and the Gateway main axis tilted by 20° with respect to the solar direction **(D)**. Note the plasma wake downstream of the station.

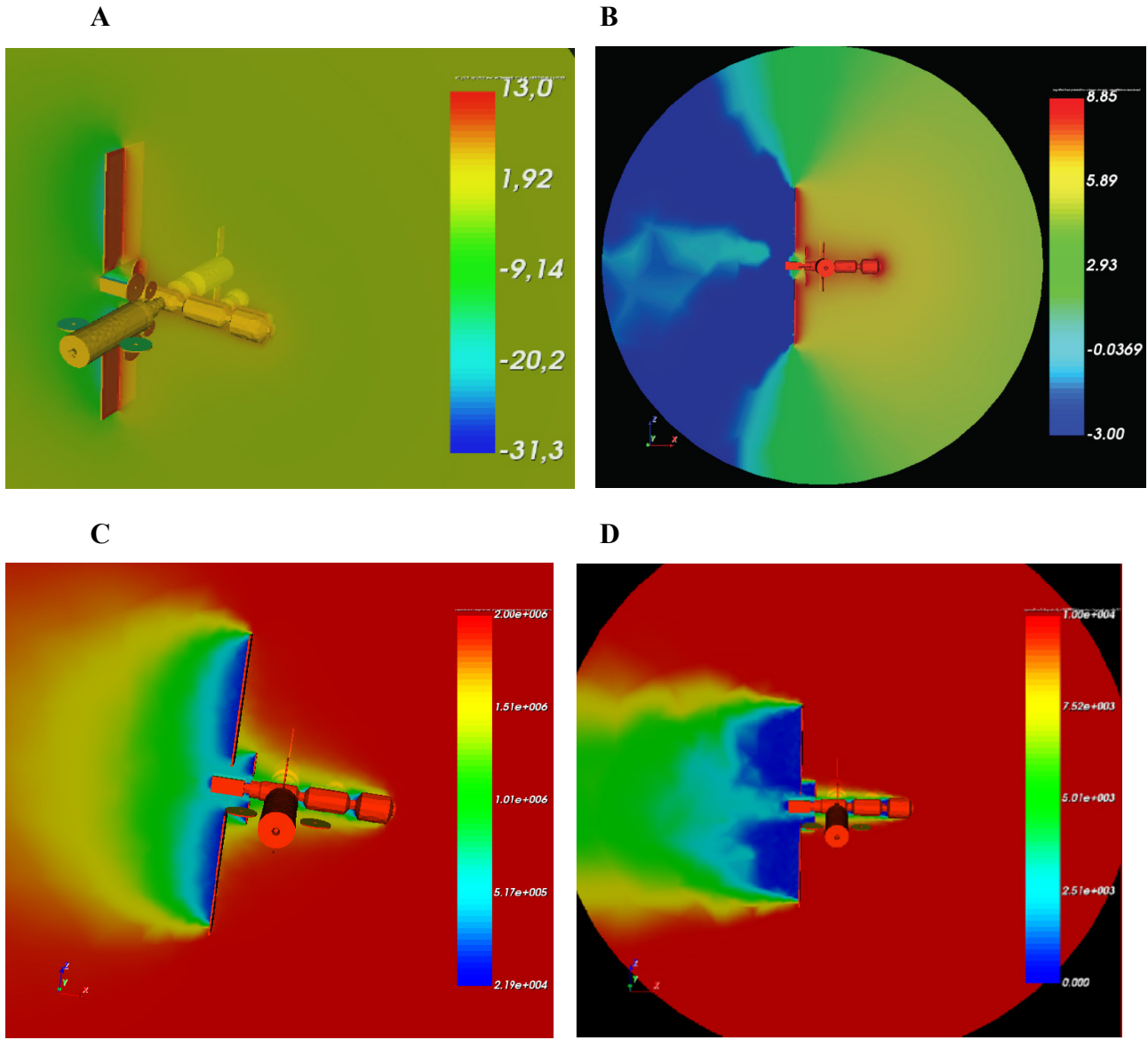

**FIGURE 14** | SPIS simulation of the volume electrostatic potential distribution (in V) in the terrestrial magnetotail **(A)**, with the Gateway aligned to the solar direction, and photoelectron density (log scale) under the same conditions **(B)**. H$^+$ ion density **(C)** and O$^+$ ion density **(D)** in the terrestrial magnetotail, both in m$^{-3}$, showing the plasma wake downstream of the station.

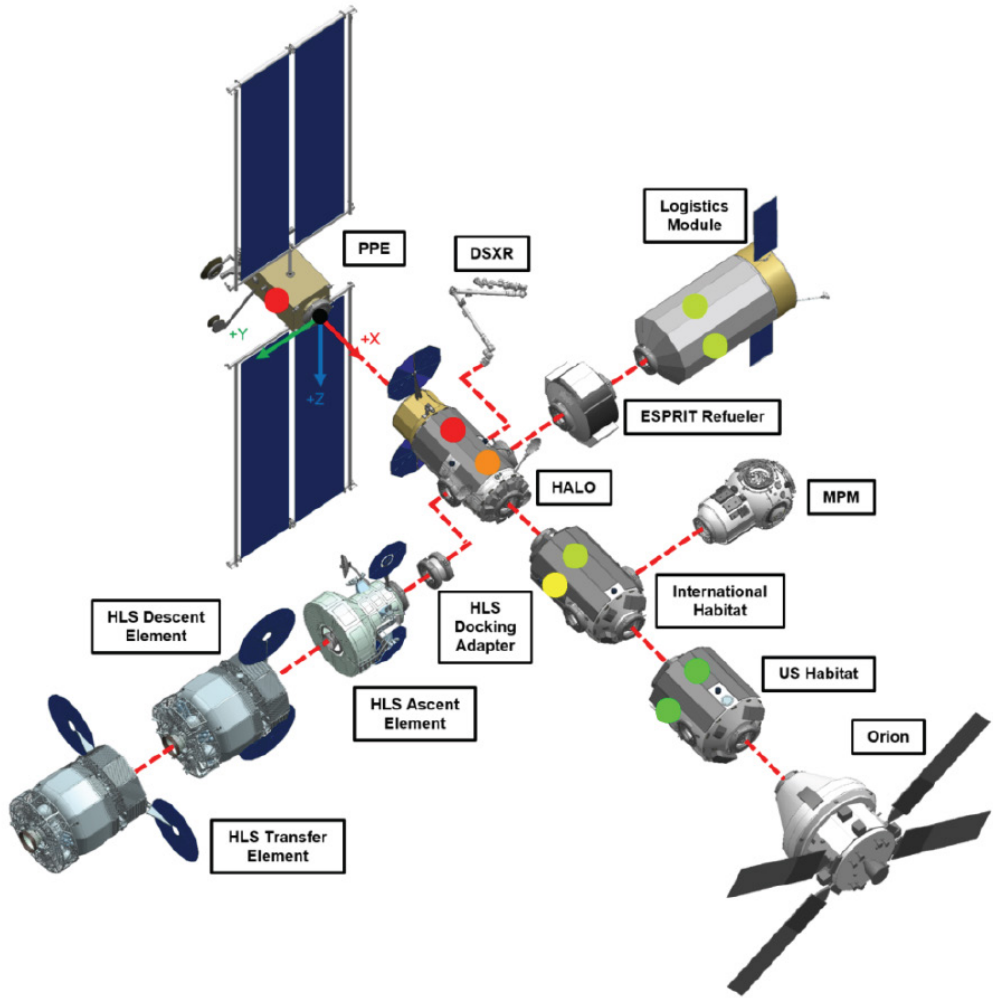

**FIGURE 15 |** Quality of the different positions on the Gateway for placing the space plasmas instruments.

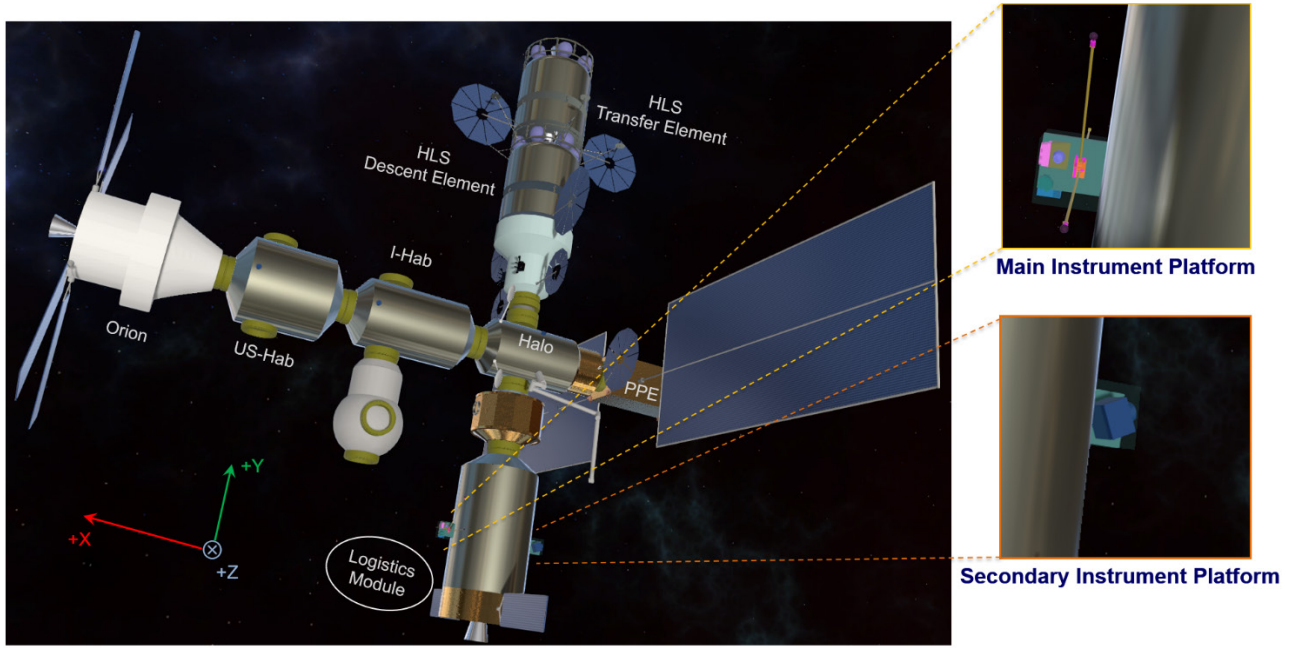

**FIGURE 16 |** Main and Secondary Instrument Platforms, mounted on the +X side and on the –X side respectively of the Logistics Module.

**A**

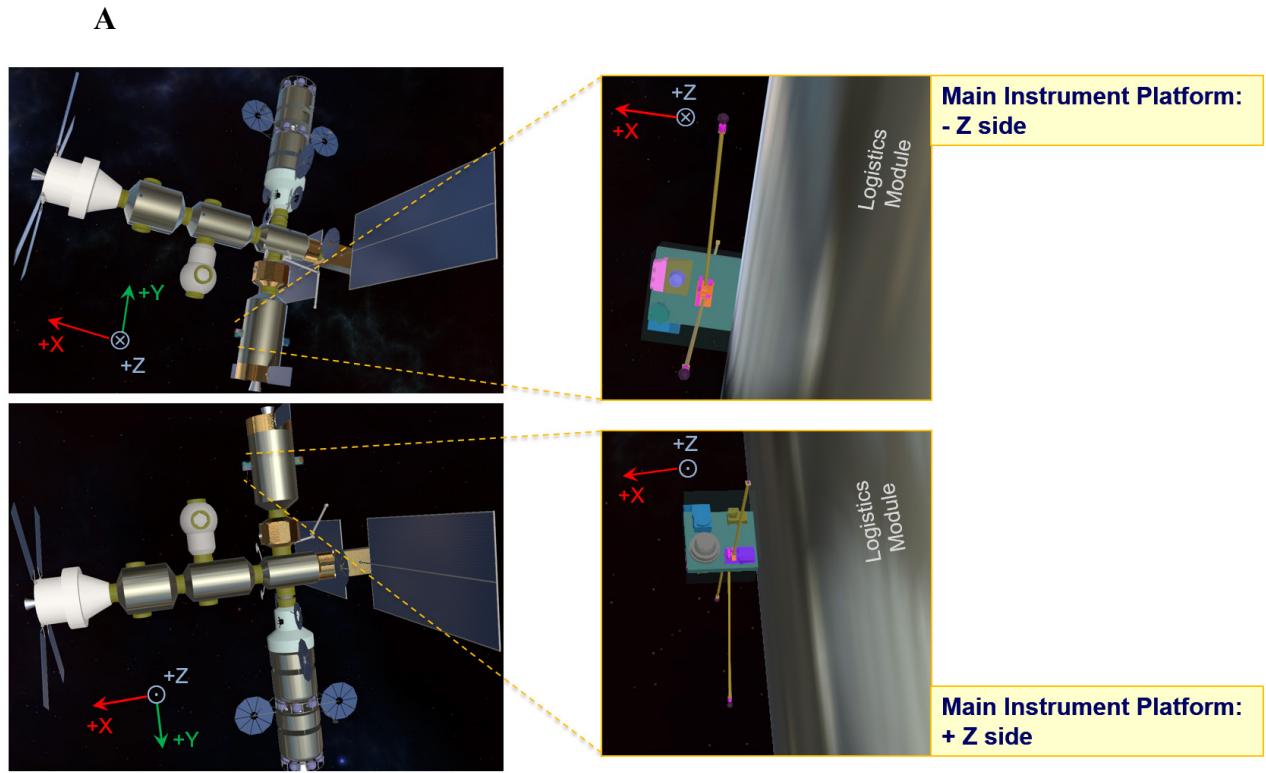

**B**

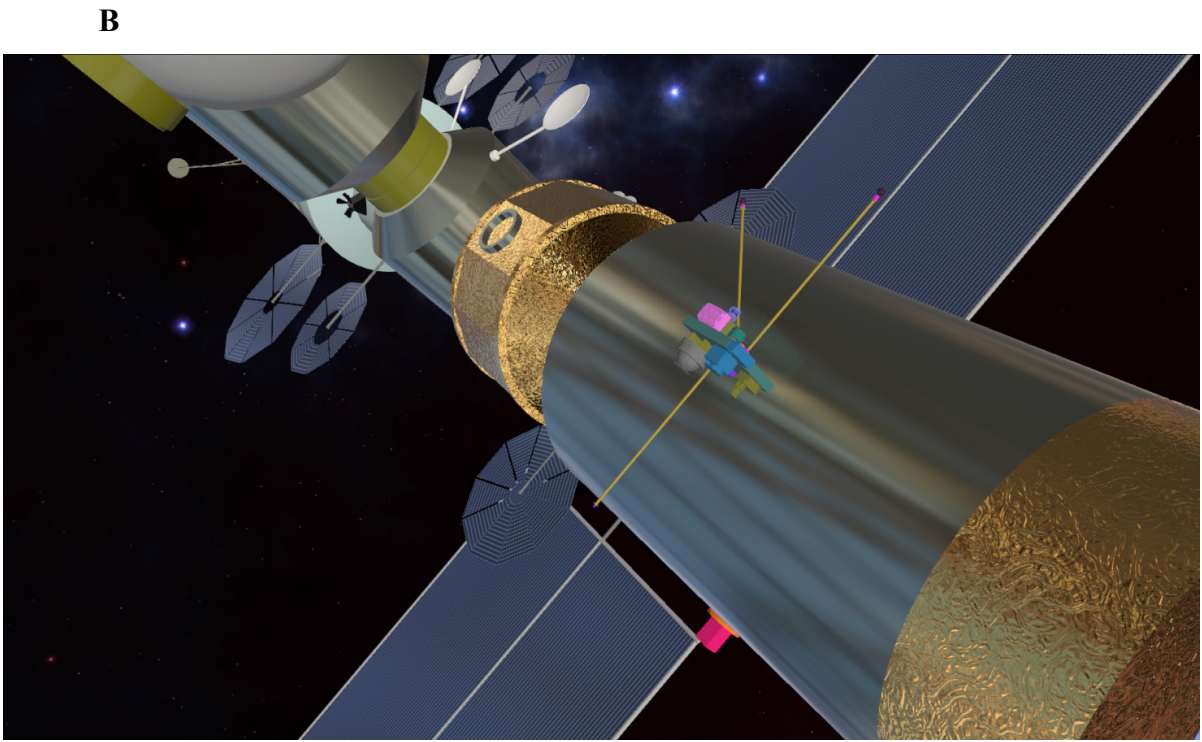

**FIGURE 17 |** The two-sided Main Instrument Platform, mounted on the Logistics Module **(A)**, and in perspective view **(B)**. The "magenta cube", on the side of the Logistics Module, is the "standalone" cGCRD instrument.

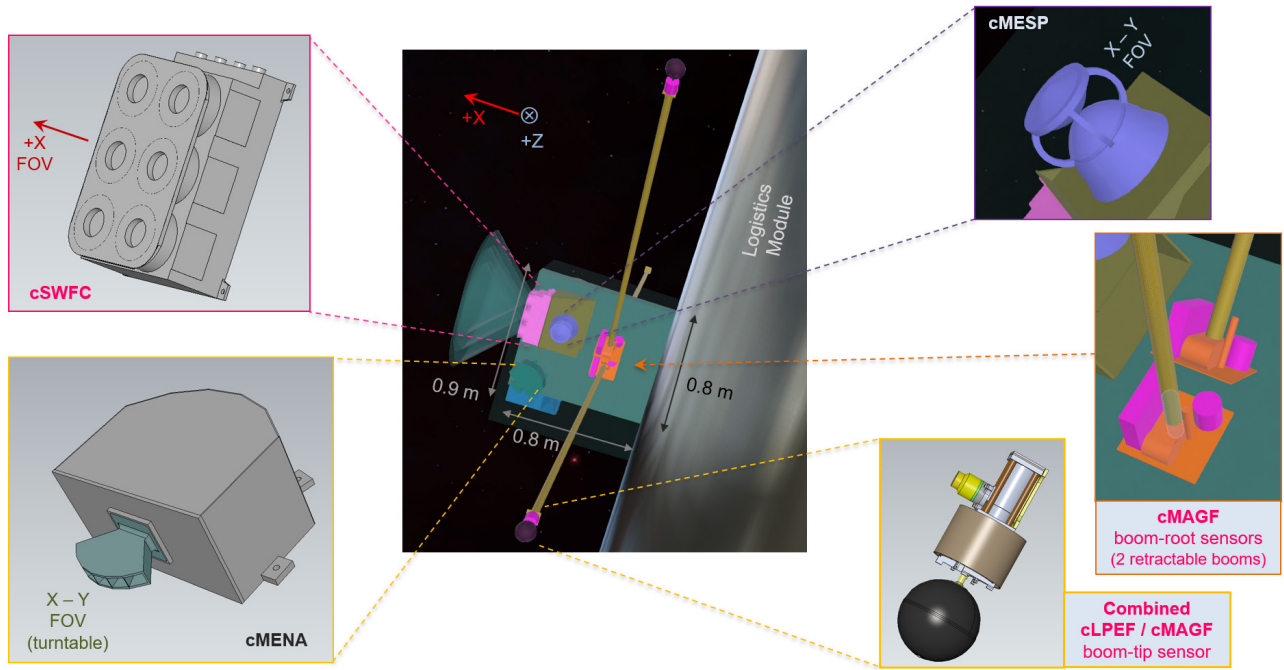

**FIGURE 18** | Main Instrument Platform –Z side, with all booms deployed.

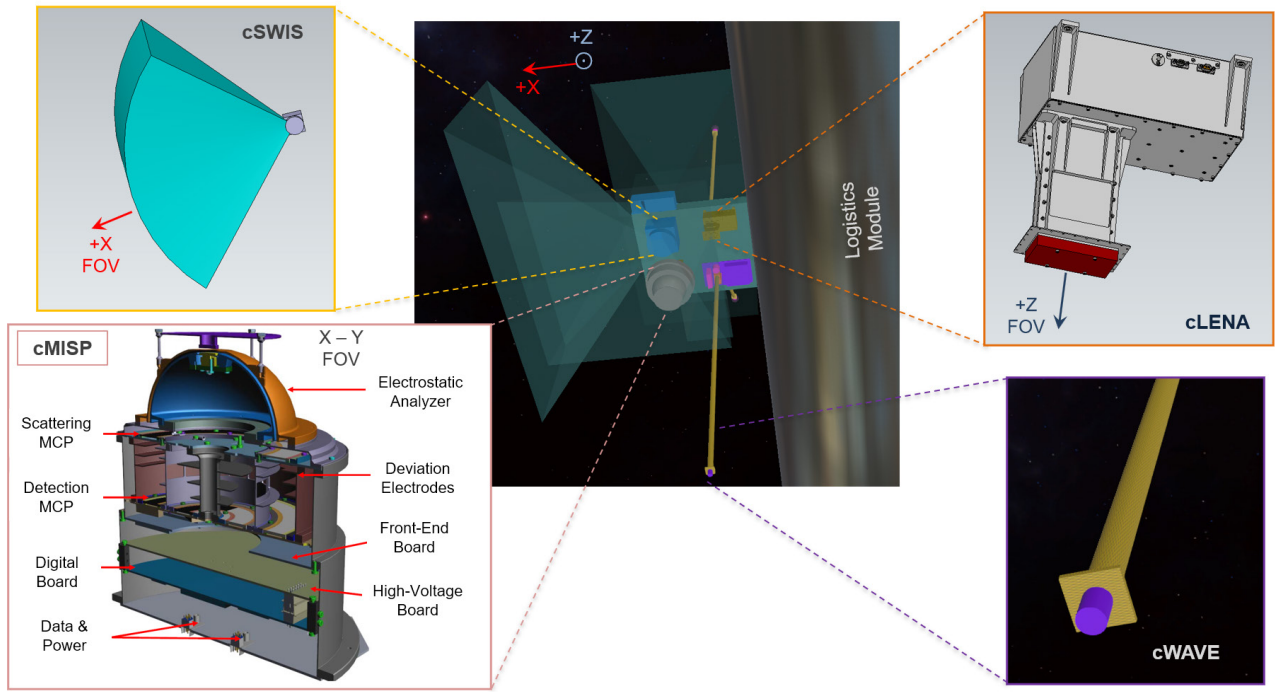

**FIGURE 19 |** Main Instrument Platform +Z side. The turquoise solid angle in the cSWIS instrument inset (upper left) represents the instrument field-of-view (FOV).

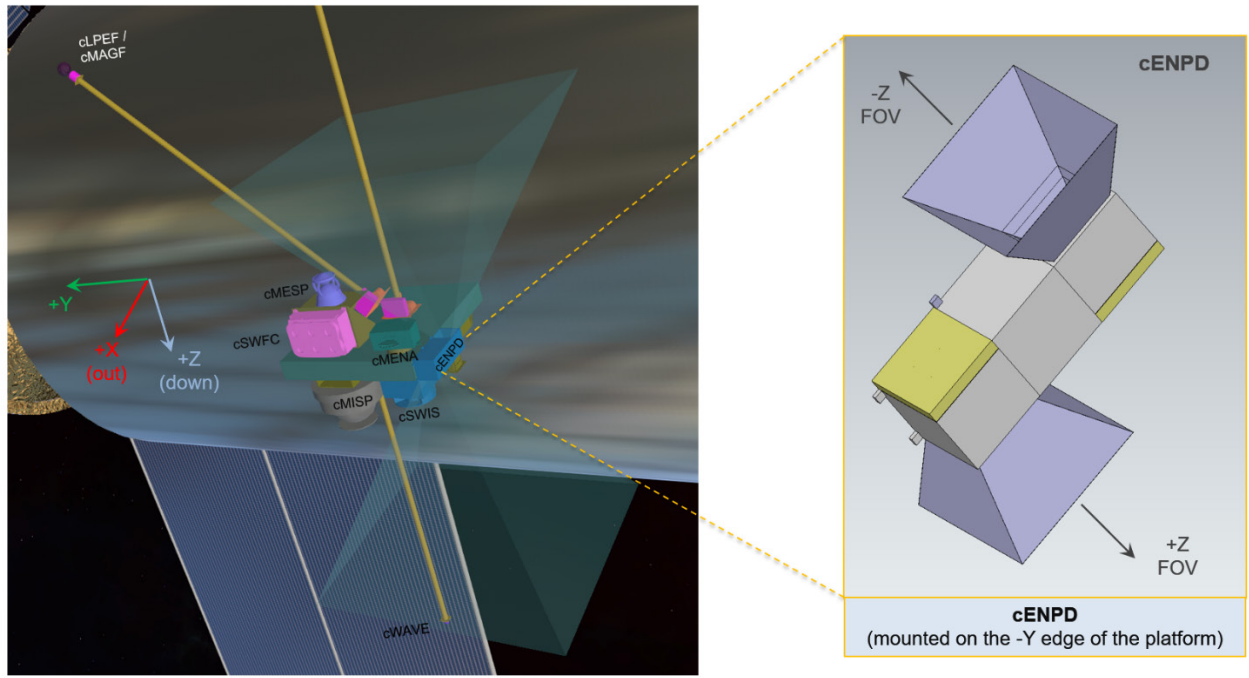

**FIGURE 20 |** Main Instrument Platform +X / −Y edges. The cENPD instrument is mounted on the edge of this platform, to get unobstructed view to both the +Z and −Z directions.

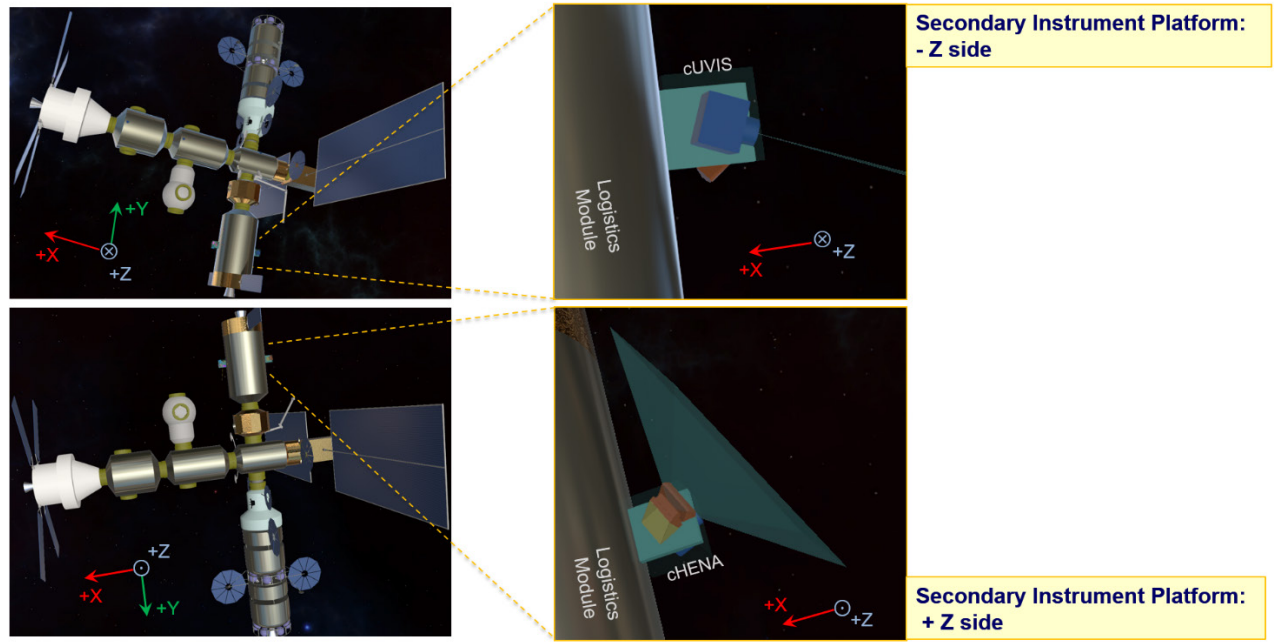

**FIGURE 21 |** The two-sided Secondary Instrument Platform, mounted on the Logistics Module.

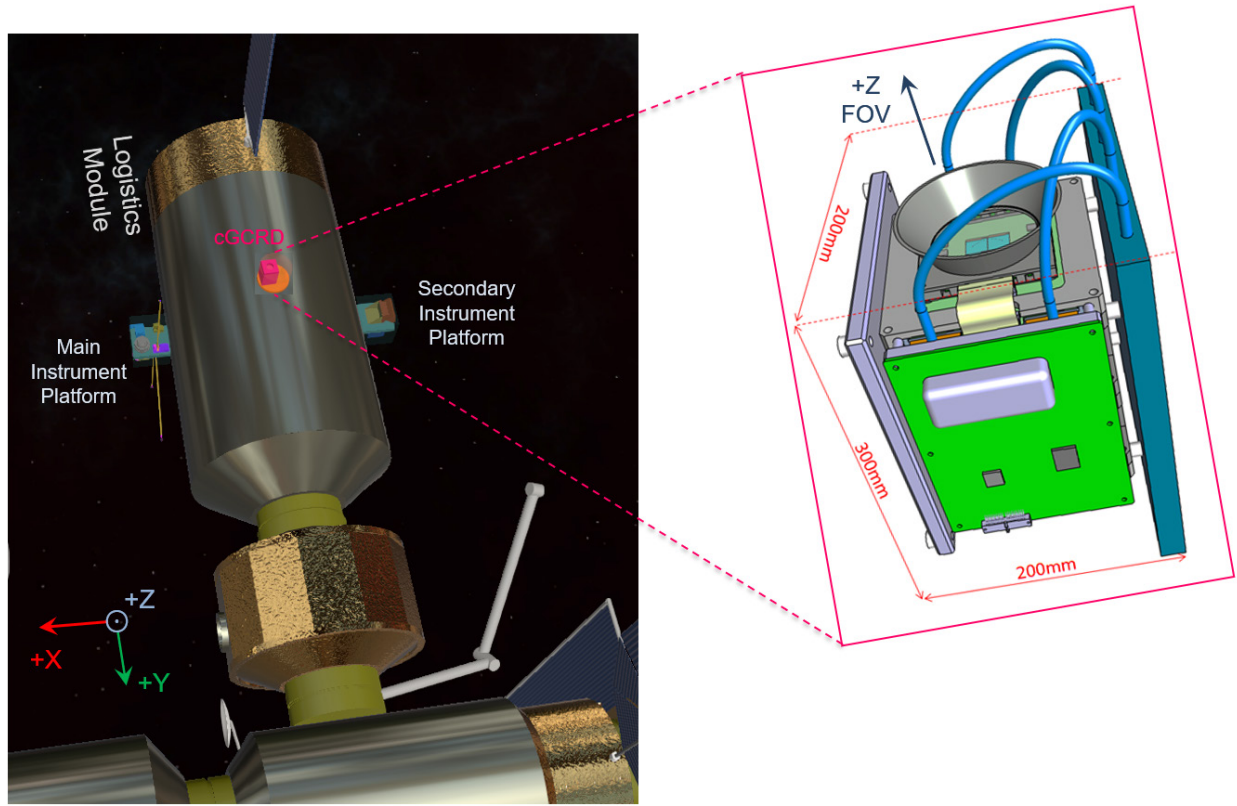

**FIGURE 22 |** cGCRD mounted as a "standalone" instrument on the Logistics Module.

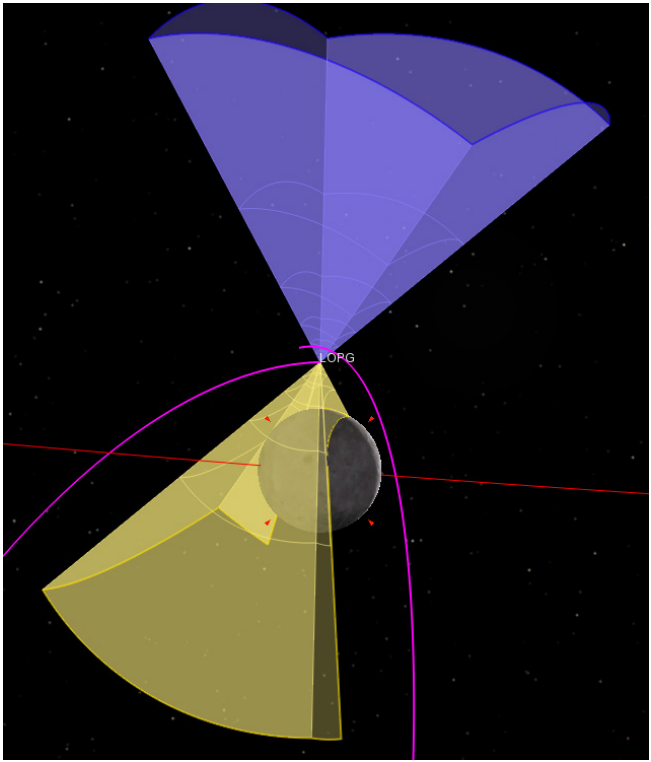

**FIGURE 23 |** cENPD instantaneous FOVs of the two oppositely directed sensor heads, near periapsis. Purple FOV: pristine energetic particles flux. Yellow FOV: Moon albedo energetic particles flux. The magenta line is the track of the center of the FOV along the Gateway orbit.

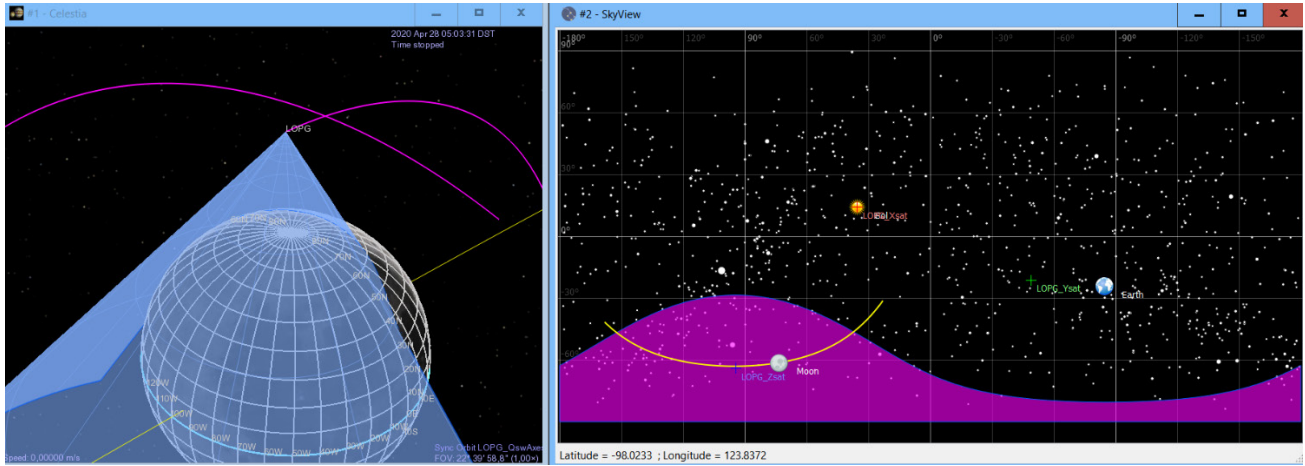

**FIGURE 24 |** cGCRD FOV. Left: cGCRD FOV near periapsis (light blue cone), dominated by the albedo GCR particles from the Moon (grid sphere). Right: projection on the sky of the cGCRD FOV along the Gateway orbit (in magenta).

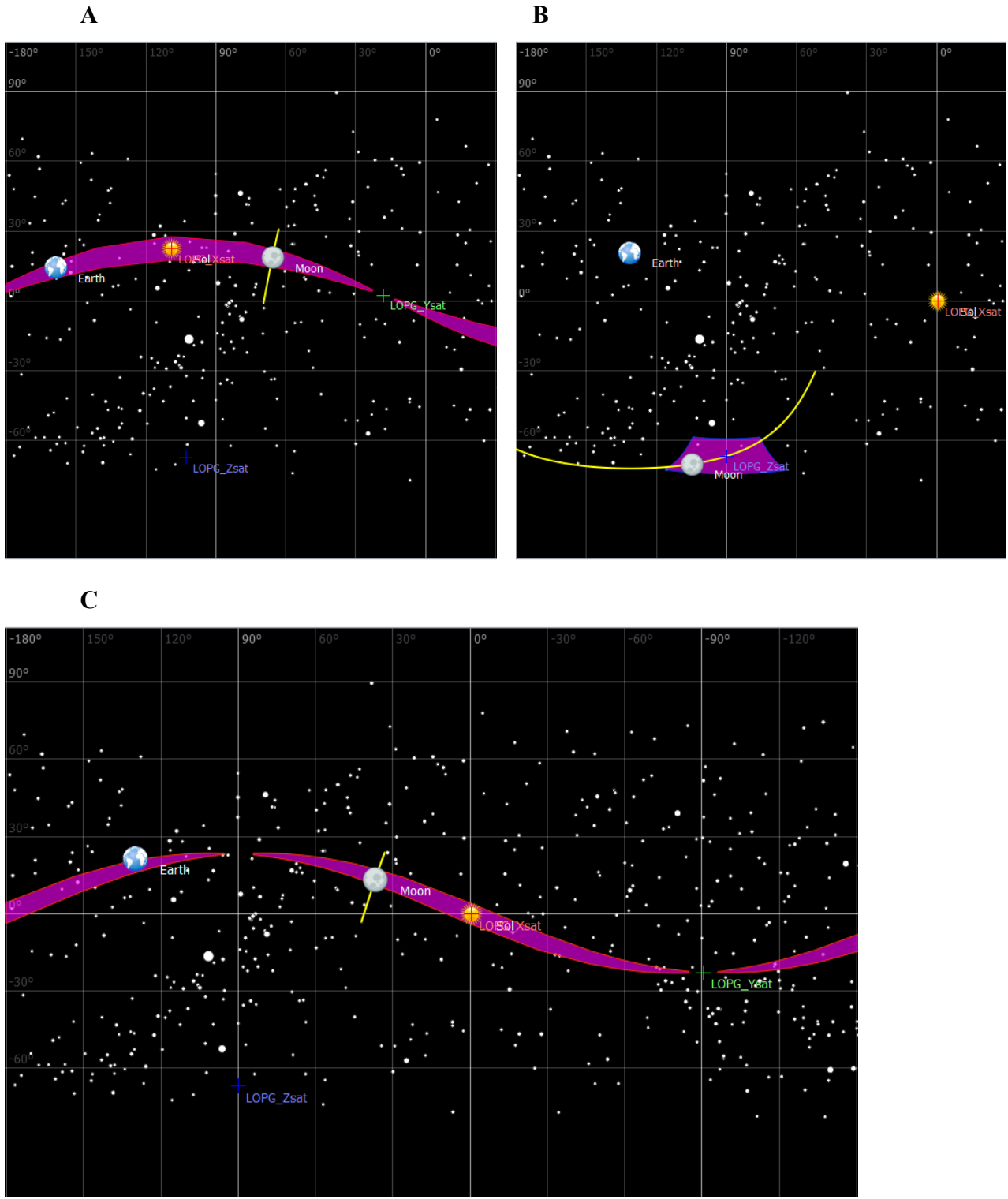

**FIGURE 25 | (A)**: Field-of-regard (total accessible FOV, taking into account the rotation of the 1-axis articulation on which the instrument is mounted) of the cMENA instrument, at a given point of the orbit. The field-of-regard (FOR), as projected on the sky, is shown in magenta. **(B)**: cLENA FOV (in magenta) near periapsis. The yellow line is the track of the center of the FOV, for the portion of the orbit close to periapsis, as projected on the sky. **(C)**: Field-of-regard of the cUVIS instrument, in magenta, as projected on the sky.